\documentclass{aa}  

\usepackage{graphicx}
\usepackage[dvipsnames]{xcolor}
\usepackage{txfonts}
\usepackage{booktabs}
\usepackage{bm}
\usepackage{ulem}
\usepackage{mathtools}
\usepackage[acronym,shortcuts]{glossaries}
\usepackage{arydshln} 
\usepackage{xspace}
\usepackage{orcidlink}
\usepackage{subfigure}
\usepackage{caption}
\usepackage{amsmath}
\usepackage{hyperref}
\hypersetup{
  colorlinks,
  citecolor=blue,
  linkcolor=blue,
  urlcolor=blue}

\newcommand{\Rsun}{\ensuremath{\,\rm{R}_{\odot}}\xspace}
\newcommand{\Rp}{\ensuremath{\,R_\mathrm{p}}\xspace}
\newcommand{\Mp}{\ensuremath{\,M_\mathrm{p}}\xspace}
\newcommand{\Mpdot}{\ensuremath{\,-\dot{M}_\mathrm{p}}\xspace}
\newcommand{\machinf}{\ensuremath{\,\mathcal{M}_\infty}\xspace}

\newcommand{\kms}{\ensuremath{\,\rm{km}\,\rm{s}^{-1}}\xspace}
\newcommand{\Msun}{\ensuremath{\,\rm{M}_{\odot}}\xspace}
\newcommand{\Msunhr}{\ensuremath{\,\rm{M}_{\odot}~\rm{h}^{-1}}\xspace}

\newcommand{\dex}{\ensuremath{\,\mathrm{dex}}\xspace}
\newcommand{\hr}{\ensuremath{\,\mathrm{h}}\xspace}

\newcommand{\yr}{\ensuremath{\,\mathrm{yr}}\xspace}

\newcommand{\gcc}{\ensuremath{\,\rm{g}\,\rm{cm}^{-3}}\xspace}

\newcommand{\K}{\ensuremath{\,\mathrm{K}}\xspace}

\newcommand{\SLH}{\texttt{SLH}\xspace }

\newacronym{KDE}{KDE}{kernel density estimate}

\allowdisplaybreaks

\begin{document}

\title{Continuous mass ablation of planets engulfed in stellar envelopes}
\titlerunning{Continuous mass ablation of planets engulfed in stellar envelopes}

\author{
  Mike Y. M. Lau \inst{1,2}\fnmsep\thanks{\email{mike.lau@h-its.org}} \orcidlink{0000-0002-6592-2036} \and
  Robert Andrassy \inst{1,2} \and
  Giovanni Leidi \inst{2} \orcidlink{0000-0001-7413-7200} \and
  Damien Gagnier \inst{1,2} \orcidlink{0000-0002-1904-2740} \and \\
  Javier Mor\'{a}n-Fraile \inst{3,4} \orcidlink{0000-0002-8918-5130} \and 
  Friedrich K. R\"{o}pke \inst{1,2,5} \orcidlink{0000-0002-4460-0097} \and
  Ilya Mandel \inst{6,7} \orcidlink{0000-0002-6134-8946}
}

\institute{
  Zentrum f{\"u}r Astronomie der Universit{\"a}t Heidelberg, Astronomisches Rechen-Institut, M{\"o}nchhofstr.\ 12-14, 69120 Heidelberg, Germany \and
  Heidelberger Institut f\"{u}r Theoretische Studien, Schloss-Wolfsbrunnenweg 35, 69118 Heidelberg, Germany \and
  Institute of Astronomy, KU Leuven, Celestijnenlaan 200D, box 2401, 3001 Leuven, Belgium \and
  Leuven Gravity Institute, KU Leuven, Celestijnenlaan 200D, box 2415, 3001 Leuven, Belgium \and
  Zentrum f\"ur Astronomie der Universit\"at Heidelberg, Institut f\"ur Theoretische Astrophysik, Philosophenweg 12, 69120 Heidelberg, Germany \and
  School of Physics and Astronomy, Monash University, Clayton, VIC 3800, Australia \and
  OzGrav: The ARC Centre of Excellence for Gravitational Wave Discovery, Clayton, VIC 3800, Australia
}

\date{Received 12 May 2026 / Accepted 06 October 2026}
 
\abstract{
  Most stars host short-period planets that are expected to be engulfed during post-main-sequence expansion. The dissolution of engulfed planets has been proposed as a possible mechanism for producing stars enriched in lithium and refractory elements. We perform three-dimensional hydrodynamical simulations of a Jupiter-like planet engulfed within a stellar envelope using the \texttt{Seven-League Hydro} code. Unlike previous studies that represent the planet as a point mass or rigid sphere, we adopted a wind-tunnel setup that resolves the planet's gaseous structure. We find that a continuous mass-ablation process operates during planetary engulfment, contrary to the common assumption that destruction occurs at a specific depth due to ram pressure, tidal forces, or thermal evaporation. The ablation rate scales nearly linearly with the wind momentum flux and is largely insensitive to the Mach number, consistent with an analytical model based on Kelvin--Helmholtz instability developing at the planetary surface. Our study provides a framework for large parameter-space studies of planetary engulfment using parametrised drag and mass-loss models. We define efficiency coefficients for drag and ablation, finding pressure-drag coefficients of 0.44--0.56 and ablation efficiencies of 0.054--0.11. Applying these coefficients to a numerically integrated inspiral through a stellar profile, we find that continuous ablation could deposit a significant fraction of the planet's mass within the convective envelope, producing observable lithium enrichment at the stellar surface. This result suggests that chemical enrichment occurs over a broader range of stellar parameters than previously thought.
}

\keywords{hydrodynamics -- methods: numerical -- planet-star interactions -- stars: chemically peculiar -- turbulence}

\maketitle
%

\section{Introduction}
\label{sec:intro}
Observational surveys have shown that the majority of stars host planets with short orbital periods \citep[e.g.][and references therein]{Mayor+11,Howard+12a,Kaminski+25}. Planets on sufficiently close orbits are ultimately expected to be engulfed as their host stars evolve, driven by tidal orbital decay and stellar expansion. Direct observational evidence of such engulfment events remains rare. The transient event ZTF SLRN-2020, a so-called subluminous red nova, has recently been identified as a possible planet-star merger involving a Neptune- or Jupiter-mass planet and a low-mass K-type main-sequence star \citep{De+23,RLau+25}. Otherwise, only indirect constraints on planetary engulfment are available through long-lived signatures imprinted on the host star, such as enhanced rotation \citep[e.g.][]{Livio+Soker02,Carlberg+09,Privitera+16b} or chemical anomalies.

Planetary engulfment is a possible pathway towards producing stars enriched in lithium and refractory elements \citep[e.g.][]{Alexander67,Aguilera-Gomez+16}. In most stars, lithium is depleted over the course of stellar evolution through proton capture at temperatures of $\,{\sim}\, 2.5\,{\times}\,10^6\K$ \citep{Burbidge+57,Bodenheimer65}, beginning during pre-main-sequence evolution. Surface lithium is further diluted during the first dredge-up, when the convective envelope deepens and mixes the stellar surface with lithium-depleted layers. The origin of lithium-rich giants ($A(\mathrm{Li})\,{>}\,1.5\dex$\footnote{The relative lithium abundance is defined as $\mathrm{A}(\mathrm{Li}) \,{=}\, 12 + \log[N(\mathrm{Li})/N(\mathrm{H})]$, where $N(\mathrm{Li})$ and $N(\mathrm{H})$ are the number densities of lithium and hydrogen atoms, respectively.}), which constitute $1\%$ of giant stars, remains uncertain \citep[e.g.][]{Brown+89,Casey+16,Gao+19}.

It is therefore crucial to understand the physical processes that cause planet mass loss inside a stellar envelope. A commonly invoked mechanism is thermal dissolution once the ambient temperature exceeds the planet's virial temperature \citep{Livio+Soker84,Siess&Livio99a,Privitera+16b}. However, this mechanism is only effective if energy can be transported into the planet on a sufficiently short timescale. This condition is not met in the engulfment of Jupiter-like planets, where the inspiral occurs over only a few orbits \citep{Lau+25}. \cite{Jia+Spruit18} instead proposed that a planet undergoes catastrophic disruption once the incident ram pressure becomes sufficiently strong. Based on this model, \cite{Soares-Furtado+21} investigated lithium enrichment from the engulfment of hot Jupiters, which are gas-giant planets with orbital periods of a few days and are found around ${\sim}\,0.1$--$1\%$ of solar-type stars \citep[e.g.][]{Mayor+11,Howard+12a}. They showed that, although stars near the main-sequence turnoff and on the subgiant branch could lead to the most statistically significant lithium enrichment, ram-pressure disruption is expected to occur only below their convective envelopes. In these cases, the disrupted planetary material cannot be mixed to the surface.

The existence of a continuous mass-ablation process would therefore greatly expand the parameter space for producing statistically significant chemical enrichment. In a previous study, \cite{Lau+25} reported gradual mass loss of a Jupiter-like planet in a three-dimensional (3D) hydrodynamical simulation that includes the entire convective envelope of an early red giant. The planet loses about 90\% of its mass in the convective envelope before reaching depths where disruption by ram pressure or by tides from the stellar core are expected, although the distribution of the removed mass is resolution dependent. They estimated that this process would lead to lithium enrichment of $\Delta A(\mathrm{Li})\approx 0.1\dex$.

These results motivate a detailed investigation of the mass-ablation mechanism. We present simulations using a wind-tunnel setup that models the local environment of an engulfed planet and represents the inspiral through the stellar envelope by an imposed background wind. This approach allows the ablation process to be studied at much higher resolution than is feasible in a global simulation. While wind-tunnel simulations of objects orbiting inside stellar envelopes have been performed previously, ablation was not modelled because the engulfed object was represented either as a point mass \citep[e.g.][]{MacLeod&Ramirez-Ruiz15,Gagnier+26,Yang+26} or as a fixed reflective sphere \citep{Thun+16,Yarza+23,Prust+Bildsten24}. Here, we instead fully resolve the internal structure of a Jupiter-like gas giant. Even in the absence of mass loss, the point-mass approximation is only adequate when gravitational drag dominates over hydrodynamical drag. This regime primarily applies to brown dwarfs and the most massive hot Jupiters, or to planets engulfed by extended red giant or asymptotic giant branch stars. Such conditions do not apply to the majority of known exoplanets \citep{Lau+25}.

Our simulations confirm the continuous mass-ablation process reported by \cite{Lau+25} and show that the mass-loss rate primarily scales with the momentum intercepted by the planet. This paper is organised as follows. In Sect.~\ref{sec:methods}, we describe the simulation setup, numerical methods, and the parameter space explored. Sect.~\ref{sec:results} presents the results, including the flow morphology (\ref{subsec:morphology}), the mass-ablation rate (\ref{subsec:ablation}) and its dependence on wind properties (\ref{subsec:ablation_mechanism}), the drag and ablation coefficients (\ref{subsec:efficiency}), and the impact of the wind Mach number (\ref{subsec:mach}). In Sect.~\ref{subsec:kelvin_helmholtz}, we derive an analytical model to explain the level of ablation measured in our simulations. Sect.~\ref{subsec:applications} demonstrates the application of the drag and ablation coefficients to a numerically integrated inspiral of an engulfed planet. Finally, Sect.~\ref{sec:conclusions} summarises our findings.

\begin{figure}
    \includegraphics[width=\linewidth]{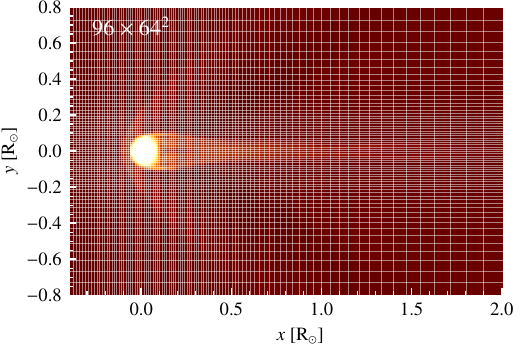}
    \caption{Illustration of the stretched grid geometry with a density cross-section from a low-resolution simulation ($96\times64^2$) using the fiducial wind parameters.}
    \label{fig:grid}
\end{figure}

\section{Methods}
\label{sec:methods}

\subsection{Setup}
\label{subsec:setup}
As in \cite{Lau+25}, we represent a hot Jupiter as an ${n\,{=}\,3/2}$ polytropic gas sphere \citep{Hubbard84,Stevenson91} with a mass of $\Mp\,{=}\,10^{-3}\Msun$ and radius of $\Rp\,{=}\,0.1\Rsun$, approximately the mass and radius of Jupiter. We neglect the heavy-element core, which comprises only a few percent of a Jovian planet's total mass and is not expected to significantly influence drag or mass ablation. The outermost layers that have lower pressure than the background wind, ranging between 3--10\% of the planet mass, represent material that is likely to have been removed at an earlier phase of the planet's inspiral, and are not mapped to the computational domain. This ensures initial pressure equilibrium between the planet's surface and the ambient medium.

We use a 3D Cartesian grid with dimensions $x\,/\Rsun\,{\in}\,[-0.4,2.0]$ and $y,z/\,\Rsun\,{\in}\,[-0.8,0.8]$. For our fiducial simulation, the grid resolution is $N_x\times N_y \times N_z\,{=}\,1536\times1024\times1024$ (1.6 billion cells). For simulations with parameter variations, the maximum resolution is $768\times 512\times512$. The effect of increasing grid resolution on the flow pattern is illustrated in Fig.~\ref{fig:res}. The Cartesian grid is stretched to increase resolution near the planet, improving the ability to preserve hydrostatic balance in the planet and to resolve local flows involved in ablation. The stretched grid geometry is illustrated in Fig.~\ref{fig:grid} using a low-resolution ($96\times64^2$) simulation snapshot. Near the planet, linear resolution is increased by a factor of two compared to a uniform grid with the same number of cells. The grid stretching functions are provided in Appendix~\ref{app:stretched_grid}.

The planet is initially centred at the origin and the background medium has uniform density, $\rho_\infty$, and velocity, $v_\infty$, along $+\hat{\mathbf{x}}$. Zero-gradient (outflow) boundaries are used except at the left $x$-boundary, which is an inlet (constant-ghost) boundary set to the wind properties. The size of the computational domain ensures that partial shock reflection at the $y$- and $z$-boundaries does not influence the planet's immediate surroundings. For example, for an inflow with Mach number $\mathcal{M}_\infty\,{=}\,2$, the reflected bow shock returns to $y\,{=}\,z\,{=}\,0$ at a distance of 28\Rp from the planet, which lies outside the computational domain. Appendix~\ref{app:domain_size} shows that doubling the domain size does not change our results.

The planet experiences a hydrodynamical drag force arising from an asymmetric pressure distribution over its surface. This drag is approximately constant because the wind density and speed are fixed at the inlet boundary. To keep the planet near the origin, where the resolution is highest, we performed the simulations in a non-inertial frame with a constant acceleration along the $\hat{\mathbf{x}}$ that approximately matches the drag. We scale the magnitude of this acceleration with the wind's momentum flux (Sect.~\ref{subsec:params}), which is later shown to be proportional to the drag acceleration (Sect.~\ref{subsec:efficiency}).

We run the simulations for $10\hr$, long after the flow reaches a quasi-steady state. In the fiducial simulation, the mass-ablation rate stabilises after ${\approx}\, 1\hr$. For the simulation with the slowest wind ($v_\infty\,{=}\,103\kms$), $10\hr$ corresponds to 53 times the wind-crossing time of the planet radius, $\Rp/v_\infty$. For comparison, the entire engulfment event simulated by \cite{Lau+25} lasts less than a day after the initial grazing phase.

During most of the inspiral, the density in the extended stellar envelope varies on scales much larger than the planet. For the case simulated by \cite{Lau+25}, the planet radius is at most one-tenth of the density scale height, except near the strongly stratified stellar surface. Therefore, we do not include any density or velocity gradient in the injected wind and neglect the gravitational field of the host star.

\subsection{Governing equations}
\label{subsec:gov-eqs}
We solve the Euler equations of gas dynamics with self-gravity in a non-inertial frame,
\begin{align}
		\label{eq:mass_conservation}
		\frac{\partial \rho}{\partial t} + \nabla \cdot (\rho \mathbf{v}) &= 0,
		\\
		\label{eq:momentum_conservation}
		\frac{\partial  (\rho \mathbf{v})}{\partial t} + \nabla \cdot  ( \rho \mathbf{v}
		\mathbf{v} + p \mathbf{I} ) &= -\rho (\nabla  \phi) - \rho \bm{a}_\mathrm{p},
		\\
		\label{eq:Energy_conservation}
		\frac{\partial (\rho e)}{\partial t} + \nabla \cdot [( \rho e + p )
		\mathbf{v}  ]  &= -\rho (\nabla \phi) \cdot \mathbf{v} - \rho \bm{a}_\mathrm{p} \cdot \mathbf{v}.
\end{align}
Here, $\rho$ is the mass density, $\mathbf{v}$ is the velocity field, $\phi$ is the gravitational potential, $\bm{a}_\mathrm{p}\,{=}\,(a_{\mathrm{p},x},0,0)$ is the frame acceleration, $e \,{=}\, e_\mathrm{int}+|\bm{v}|^2/2$ is the hydrodynamic energy per unit mass, $e_\mathrm{int}$ is the specific internal energy, and $\mathbf{I}$ is the unit tensor. These equations are closed by an equation of state, $p\,{=}\,p(\rho,e_\mathrm{int})$, and specification of the gravitational potential. In our simulations, we assume an ideal gas law,
\begin{align}
    p(\rho,e_\mathrm{int})=(\gamma-1)\rho e_\mathrm{int},
\end{align}
with an adiabatic index $\gamma$ of $5/3$. The gravitational potential is obtained from Poisson's equation, 
\begin{align}
    \label{eq:poisson}
    \nabla^2 \phi  = 4\pi G (\rho-\rho_\infty),
\end{align}
where $G$ is the gravitational constant. The background wind density, $\rho_\infty$, is subtracted from the source term so that gravity only arises from the planet and from perturbations of the background.

To monitor the planet's mass and gain insight into the ablation mechanism, we evolve a passive scalar, $\psi$, that traces planetary material and evolves according to
\begin{equation}
    \label{eq:tracer_conservation}
    \frac{\partial (\rho \psi)}{\partial t} + \nabla \cdot (\rho \psi\mathbf{v}) = 0.
\end{equation}
The passive scalar is initialised to unity inside the planet and zero in the rest of the spatial domain.

\subsection{Numerical methods}
\label{subsec:SLH}
Equations~(\ref{eq:mass_conservation})--(\ref{eq:tracer_conservation}) are solved numerically with the \texttt{Seven-League Hydro} (\SLH) code \citep[]{miczek2013a,edelmann2014a,miczek2015,edelmann2021a,leidi2024}. \SLH is a finite-volume, Godunov-like code optimised for modelling highly subsonic (magneto-)hydrodynamic phenomena in stellar interiors, such as convective boundary mixing \citep[]{horst2021a,andrassy2022,andrassy2024}, the excitation of internal gravity waves by core convection \citep[]{horst2020a}, and turbulent dynamos in late burning shells of massive stars \citep[]{leidi2022,leidi2023}. In addition to a wide variety of low-Mach-number solvers, \SLH also includes a semi-discrete adaptation of the piecewise-parabolic-method of \cite{colella1984a} and the Harten-Lax-van Leer-Contact approximate Riemann solver of \cite{toro1994} to robustly and accurately capture both shock waves and contact discontinuities. The set of conserved quantities ($\rho$, $\rho \mathbf{v}$, $\rho e$, $\rho \psi$) is marched in time in fully unsplit fashion using a semi-discrete scheme together with the third-order time-explicit strong-stability-preserving RK3 method of \cite{shu1988}. The time step is determined from the Courant-Friedrichs-Lewy criterion \citep{courant1928} with a Courant factor of 0.4.

Poisson's equation is coupled to Eq.~(\ref{eq:momentum_conservation}) and (\ref{eq:Energy_conservation}) using first-order Godunov time splitting. It is discretised using a finite-difference scheme for non-uniform Cartesian grids that is formally first-order but practically second-order for a smoothly varying mesh. The equation is solved iteratively with a matrix-free implementation of the bi-conjugate gradient stabilised (BiCGSTAB) method of \cite{vandervorst1992}. At the grid boundaries, $\phi$ is approximated at each time step by the monopole expansion of the planet's gravitational potential. Our implementation follows \cite{Leidi+26}, and further details of our implementation of self-gravity are described in Appendix~\ref{sec:gravity-solver}.

\subsection{Parameter space}
\label{subsec:params}

\begingroup
\renewcommand{\arraystretch}{1.2}
\begin{table*}[h!]
    \centering
    \caption{Simulation parameters and main results.}
    \label{tab:summary}
    \begin{tabular}{@{}lcccccccccc@{}}
        \hline\hline
        & $f_\mathrm{m}$ & $f_\mathrm{e}$ & $\frac{\rho_\infty}{\rho_{\infty,0}}$ & $\frac{v_\infty}{v_{\infty,0}}$ & $\mathcal{M}_\infty$ & $(R_\mathrm{BHL}/\Rp)^2$ & $\log\Big(\frac{\langle\Mpdot\rangle}{\Msunhr}\Big)$ & $\langle\eta_\mathrm{pres}\rangle$ & $\langle\eta_\mathrm{abl}\rangle$ & $\langle\Delta x_\mathrm{SO}\rangle$ \\
        &  &  &  &  &  & $[10^{-2}]$ & & & & $[10^{-2}\Rsun]$ \\ \midrule
        (i)$^{\ast}$      & 1     & 1     & 1     & 1     & 2 & 0.33  & $-4.78^{+0.05}_{-0.08}$ & $0.464^{+0.017}_{-0.015}$ & $0.108^{+0.010}_{-0.007}$    & $4.02^{+0.20}_{-0.12}$ \\
        (ii)$^{\ast\ast}$ & 2     & 1     & 8     & 0.5   & 2 & 5.2   & $-4.67^{+0.08}_{-0.26}$ & $0.556^{+0.032}_{-0.034}$ & \dots                        & \dots                     \\
        (iii)             & 1     & 2     & 0.25  & 2     & 2 & 0.020 & $-5.02^{+0.03}_{-0.06}$ & $0.444^{+0.010}_{-0.009}$ & $0.0598^{+0.0046}_{-0.0051}$ & $3.95^{+0.11}_{-0.10}$ \\
        (iv)$^{\ast\ast}$ & 3/2   & 2/3   & 7.59  & 0.44 & 2 & 8.4   & $-4.79^{+0.10}_{-0.21}$ & $0.541^{+0.048}_{-0.054}$  & \dots                        & \dots                     \\
        (v)               & 2/3   & 3/2   & 0.132 & 2.25  & 2 & 0.013 & $-5.21^{+0.03}_{-0.06}$ & $0.435^{+0.010}_{-0.011}$ & $0.0570^{+0.0035}_{-0.0041}$ & $4.08^{+0.05}_{-0.13}$ \\
        (vi)$^{\ast\ast}$ & 1     & 1/2   & 4     & 0.5   & 2 & 5.2   & $-4.83^{+0.02}_{-0.03}$ & $0.526^{+0.032}_{-0.023}$ & \dots                        & \dots                     \\
        (vii)             & 1/2   & 1     & 0.125 & 2     & 2 & 0.020 & $-5.33^{+0.04}_{-0.05}$ & $0.436^{+0.011}_{-0.009}$ & $0.0545^{+0.0029}_{-0.0034}$ & $4.09^{+0.10}_{-0.11}$ \\ \hdashline
        (viii)            & 1     & 1     & 1     & 1     & 3 & 0.41  & $-4.84^{+0.09}_{-0.10}$ & $0.475^{+0.010}_{-0.009}$ & $0.0736^{+0.015}_{-0.015}$   & $2.77^{+0.12}_{-0.11}$ \\
        (ix)              & 1     & 1     & 1     & 1     & 4 & 0.45  & $-4.80^{+0.09}_{-0.12}$ & $0.469^{+0.011}_{-0.013}$ & $0.0740^{+0.018}_{-0.011}$   & $2.51^{+0.18}_{-0.18}$ \\
        (x)               & 1     & 1     & 1     & 1     & 5 & 0.47  & $-4.83^{+0.12}_{-0.12}$ & $0.465^{+0.009}_{-0.013}$ & $0.0700^{+0.015}_{-0.015}$   & $2.33^{+0.16}_{-0.12}$ \\
        \bottomrule
    \end{tabular}
    
    \tablefoot{Each simulation is fully described by the set of parameters, $(\mathcal{M}_\infty,f_\mathrm{m},f_\mathrm{e})$ or $(\mathcal{M}_\infty,\rho_\infty,v_\infty)$. In the table headers, $\rho_\infty$ is the wind density, $v_\infty$ is the wind speed, $\mathcal{M}_\infty$ is the wind Mach number, $f_\mathrm{m}$ is the normalised wind momentum flux (Eq. (\ref{eq:fm})), $f_\mathrm{e}$ is the normalised wind energy flux (Eq. (\ref{eq:fe})), and $\Rp/R_\mathrm{BHL}$ is the ratio of the planet radius to the Bondi--Hoyle--Lyttleton radius (Eq. (\ref{eq:BHL})) as a measure of the relative magnitudes of pressure and gravitational drag. The last four columns report the quasi-steady mass-ablation rate, pressure-drag coefficient, ablation coefficient, and shock stand-off distance. The error bars reflect stochasticity in their measured values and correspond to $1\sigma$ confidence intervals.\\
        \tablefoottext{$\ast$}{Also performed for an adiabatic index of $\gamma=7/5$ at the grid resolution of $768\times512^2$.}
        \tablefoottext{$\ast\ast$}{Performed in an inertial frame with a uniform Cartesian grid (Sect.~\ref{subsec:params}). We do not report $\langle\eta_\mathrm{abl}\rangle$ and $\langle\Delta x_\mathrm{SO}\rangle$ for these cases because they do not reach a quasi-steady value.}
    }
\end{table*}
\endgroup

For a given planet structure, our simulations can be fully described by three independent parameters: the wind speed, $v_\infty$; the wind density, $\rho_\infty$; and the wind Mach number, $\mathcal{M}_\infty\,{=}\,v_\infty/c_{s,\infty}$, where $c_{s,\infty}$ is the wind sound speed. The parameters characterising our fiducial simulation ($\rho_{\infty,0}$, $v_{\infty,0}$, $\mathcal{M}_{\infty,0}$) are taken from the 3D simulation of hot Jupiter engulfment performed by \cite{Lau+25}. Unlike our wind-tunnel setup, their simulation includes the entire convective envelope of the 4\Rsun red giant as well as part of the radiative hydrogen-burning shell beneath it. We adopted the stellar density, sound speed, and azimuthal velocity from their model when the hot Jupiter has spiralled inwards to a separation of 2.8\Rsun from the stellar centre, yielding $\rho_{\infty,0}\,{=}\,0.0118\gcc$, $v_{\infty,0}\,{=}\,231\kms$, and $\mathcal{M}_{\infty,0}\,{=}\,2$.

To more closely connect our simulations to potential ablation processes, it is convenient to parametrise $\rho_\infty$ and $v_\infty$ in terms of the wind's energy flux, $\rho_\infty v_\infty^3$, and momentum flux, $\rho_\infty v_\infty^2$. For our fiducial parameters, these evaluate to $\rho_{\infty,0} v_{\infty,0}^2\,{=}\,6.30\times10^{12}~\rm{g}~\rm{cm}^{-1}~\rm{s}^{-2}$ and $\rho_{\infty,0} v_{\infty,0}^3\,{=}\,1.45\times10^{20}~\rm{g}~\rm{s}^{-3}$. In a series of simulations listed in Table~\ref{tab:summary}, we systematically explore variations about this fiducial case. We define the dimensionless parameters $f_\mathrm{m}$ and $f_\mathrm{e}$ as the wind momentum and energy fluxes normalised by their fiducial values, i.e.,
\begin{align}
    f_\mathrm{m} &:= \rho_\infty v_\infty^2 / (\rho_{\infty,0} v_{\infty,0}^2),  \label{eq:fm} \\
    f_\mathrm{e} &:= \rho_\infty v_\infty^3 / (\rho_{\infty,0} v_{\infty,0}^3).  \label{eq:fe}
\end{align}
The set $(\mathcal{M}_\infty, f_\mathrm{m}, f_\mathrm{e})$ therefore provides another parametrisation of a simulation. The parameter variations beyond the fiducial case, however, do not necessarily correspond to realistic conditions encountered during planetary engulfment. Instead, they are chosen with the goal of uncovering the physical mechanisms governing planet ablation and drag. 

Simulations (ii)--(vii) represent variations in the wind energy and momentum fluxes while keeping the Mach number fixed. This is achieved by adjusting the wind pressure, $p_\infty$, which determines the sound speed through $c_{s,\infty}^2\,{=}\,\gamma p_\infty/\rho_\infty$. It is also possible to instead fix $p_\infty$, but we chose to hold $\mathcal{M}_\infty$ constant in order to preserve the basic flow geometry, including the bow-shock opening angle and the shock stand-off distance (Sect.~\ref{subsec:mach}). Simulation (ii) doubles the momentum flux while simulation (iii) doubles the energy flux, keeping the other flux constant. Simulations (iv) and (v) simultaneously vary the momentum and energy fluxes in opposite directions. Case (iv) increases $f_\mathrm{m}$ and decreases $f_\mathrm{e}$ by a factor of 1.5, while case (v) has the opposite variation, decreasing $f_\mathrm{m}$ and increasing $f_\mathrm{e}$ by the same factor. These two cases allow us to assess whether planet ablation correlates more strongly with momentum or energy injection. Simulations (viii)--(x) represent variations in the Mach number while keeping $f_\mathrm{m}\,{=}\,f_\mathrm{e}\,{=}\,1$, which is also achieved by adjusting the wind pressure.

As we will show in Sect.~\ref{sec:results}, the pressure drag acting on the planet is proportional to the wind momentum flux. The frame acceleration, $\bm{a}_\mathrm{p}$, must therefore scale linearly with $f_\mathrm{m}$ in order to keep the planet stationary within the computational domain. We obtained satisfactory results with $a_{\mathrm{p},x}\,{=}\, (180~\mathrm{cm}~\mathrm{s}^{-2})f_\mathrm{m}$, where the residual planet displacement at the end of the fiducial simulation is only $\approx 20\%$ of the planet radius.

In addition to pressure drag, the planet also experiences gravitational drag, or dynamical friction \citep{Chandrasekhar43,Ostriker1999}, due to gravitational attraction to its overdense wake and Mach cone. Unlike the simulations of \cite{Yarza+23} and \cite{Prust+Bildsten24}, gravitational drag in our simulations arises self-consistently from the inclusion of self-gravity rather than from a fixed gravitational potential. However, in all simulations, we estimate its contribution to be initially negligible or subdominant compared to pressure drag. This is expected for Jovian or lower-mass planets engulfed by stars with radii $\lesssim 10^2\Rsun$ \citep{Lau+25}. Table~\ref{tab:summary} lists the squared ratio of the Bondi--Hoyle--Lyttleton radius to the planet radius, $(R_\mathrm{BHL}/\Rp)^2$, which approximately provides the ratio between gravitational and pressure drag. We adopted the expression $R_\mathrm{BHL}\,{=}\,2G\Mp/(v_\infty^2+c_{s,\infty}^2)$ \citep{Hoyle-Lyttleton1939,Bondi1952}. In terms of the simulation parameters $(f_\mathrm{m},f_\mathrm{e},\mathcal{M}_\infty)$, we obtain
\begin{align}
    \frac{R_\mathrm{BHL}}{\Rp} &= \Bigg(\frac{f_\mathrm{m}}{f_\mathrm{e}}\Bigg)^2 \frac{2GM_\mathrm{p}}{v_{\infty,0}^2 R_\mathrm{p} (1+\mathcal{M}_\infty^{-2})}.
    \label{eq:BHL}
\end{align}
With our fiducial parameters, $(R_\mathrm{BHL}/\Rp)^2\,{=}\,0.0033 (f_\mathrm{m}/f_\mathrm{e})^4$, meaning gravitational drag is only ${\sim}\,0.1\%$ of the pressure drag. Even in cases with denser and slower winds, this ratio increases to at most ${\sim}10$\%. Despite this estimate, we find that gravitational drag becomes significant in the three ``dense-wind'' simulations where $f_\mathrm{m}\,{>}\,f_\mathrm{e}$ (marked with double asterisks in Table~\ref{tab:summary}). In these runs, the drag is dominated by a large-scale density enhancement behind the bow shock, and grows steadily prior to gravitational collapse (Sect.~\ref{subsec:ablation_mechanism}). This triggers numerical instabilities when subsonic flows, and eventually inflows, develop near the $y$- and $z$-boundaries. Because a uniform frame acceleration cannot compensate for the growing gravitational drag and hold the planet still, we instead perform the dense-wind simulations on a uniform grid in the inertial frame. This reduces the linear resolution at the planet by a factor of two compared to the stretched grid with the same number of cells. Nevertheless, we show that the measured mass ablation rates are converged or very close to convergence with resolution.

\section{Results}
\label{sec:results}

\begin{figure}
    \centering
    \includegraphics[width=\columnwidth]{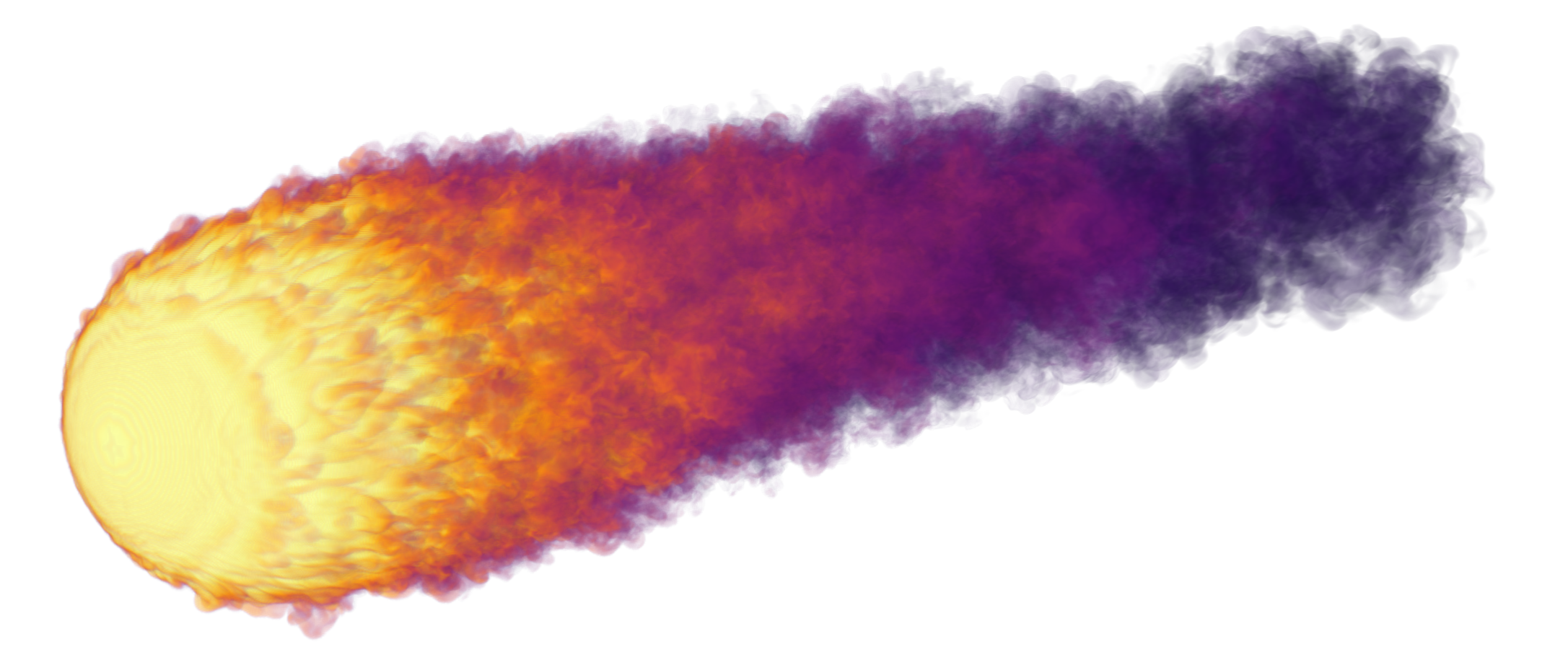}
    \caption{Volume rendering showing planet ablation in the fiducial simulation at a grid resolution of $1536\times 1024^2$. Planetary material is traced by the passive scalar, $\psi$. An animated version of this figure is available via a \href{https://youtube.com/playlist?list=PLZvLCitllq5s5MXv1hH1TnzruyMMslkZ9&si=KW-R2q9AHpAuUZuc}{YouTube playlist}.}
    \label{fig:3d_render}
\end{figure}

\begin{figure}
    \centering
    \includegraphics[width=\linewidth]{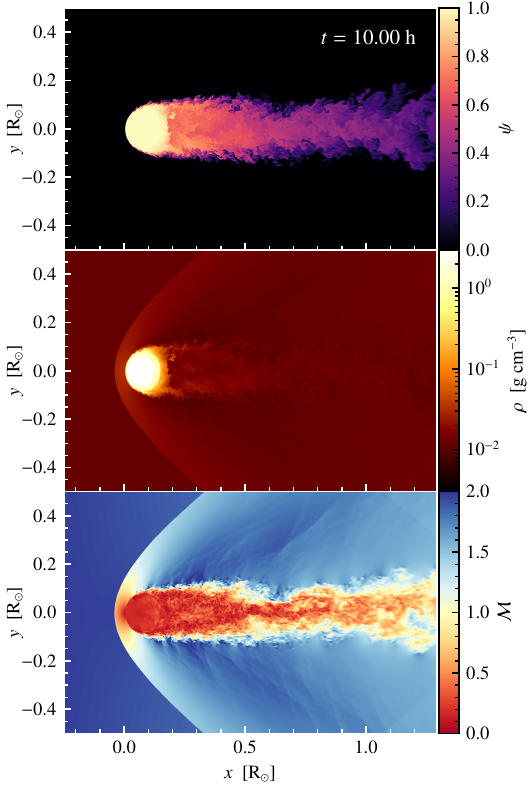}
    \caption{Slices of the fiducial simulation in the $z\,{=}\,0$ plane at the last snapshot ($t\,{=}\,10\hr$). Top panel: Passive scalar, $\psi$, which shows planetary material. Middle panel: Density, $\rho$. Bottom panel: Mach number, $\mathcal{M} = |\mathbf{v}|/c_s$. An animated version of this figure is available via a \href{https://youtube.com/playlist?list=PLZvLCitllq5s5MXv1hH1TnzruyMMslkZ9&si=KW-R2q9AHpAuUZuc}{YouTube playlist}.}
    \label{fig:psi_rho_mach}
\end{figure}

Figure~\ref{fig:3d_render} shows a volume rendering of the planetary material in the fiducial simulation ($f_\mathrm{m}\,{=}\,f_\mathrm{e}\,{=}\,1$), while Fig.~\ref{fig:psi_rho_mach} shows slices in the $xy$-plane long after the flow has reached a quasi-steady state. The top panel and the volume rendering show Kelvin--Helmholtz vortex sheets developing on the planet's surface, which are swept downstream and mixed into the turbulent wake. This likely contributes to mass ablation, and we quantify the rate of mass loss in Sect.~\ref{subsec:ablation}. A prominent bow shock forms upstream of the planet and can be seen in the middle and bottom panels of Fig.~\ref{fig:psi_rho_mach}. The bow shock is associated with a discontinuous drop in Mach number and a jump in density. Close to the stagnation streamline, where the shock is strongest and normal to the flow, these changes are in close agreement with the Rankine--Hugoniot jump conditions, which predict a post-shock Mach number of $0.61$ and a factor of $2.3$ jump in density, for $\gamma=5/3$ and $\mathcal{M}_\infty\,{=}\,2$. Farther away from the planet, the bow shock becomes weaker and asymptotically approaches a Mach cone with a half-opening angle given by $\arcsin(\mathcal{M}_\infty^{-1})$ \citep[e.g.][]{Landau+Lifshitz59}, which is $30^\circ$ in this case.

\subsection{Shock morphology}
\label{subsec:morphology}

\begin{figure*}
    \centering
    \includegraphics[width=\textwidth]{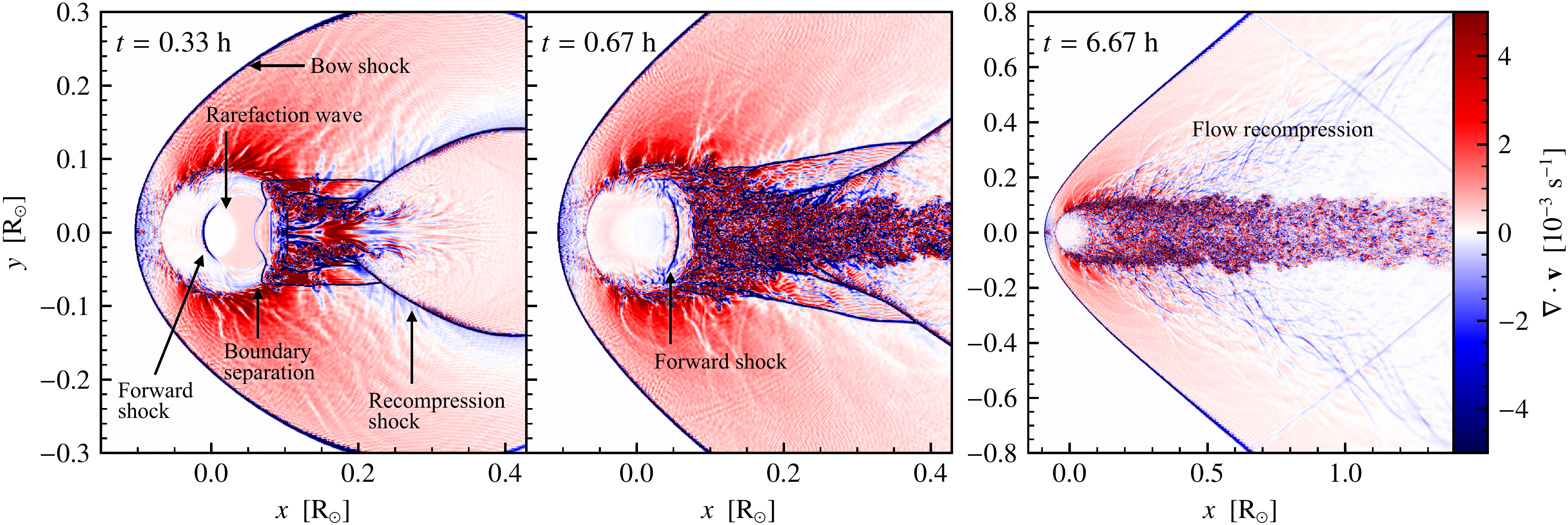}
    \caption{Velocity divergence in the $z=0$ slice of the fiducial simulation. Shock fronts appear as thin blue filaments of strongly negative velocity divergence, indicating abrupt compression. The sequence of panels shows the formation and evolution of the bow shock, the forward shock, and the recompression shock. The rightmost panel uses an extended plotting range, showing the downstream recompression of the flow. Animated versions of this figure are available via a \href{https://youtube.com/playlist?list=PLZvLCitllq5s5MXv1hH1TnzruyMMslkZ9&si=KW-R2q9AHpAuUZuc}{YouTube playlist}.}
    \label{fig:shock_sweeping}
\end{figure*}

To identify shocks and compression/rarefaction regions, we plot slices of velocity divergence in Fig.~\ref{fig:shock_sweeping}. Shocks appear as thin blue lines, indicating a step-like decrease in velocity, while rarefaction regions have positive velocity divergence and are shown in red. At the start of the simulation, the background velocity is set to be that of the wind, launching a forward shock that sweeps through the planet and a reverse shock propagating in the opposite direction that forms the standing bow shock (left panel). Boundary layer separation occurs past the halfway point along the planet's surface, at $x\,{\approx}\,0.07\Rsun$. The separated flow immediately behind the planet undergoes strong expansion and develops into a turbulent wake, divided from the laminar region by a separation shock. Further downstream, near $x=0.2\Rsun$ in the left panel, a recompression shock increases the pressure of the flow to match that in the wake region. This flow morphology is consistent with the regime of pressure-dominated drag ($R_\mathrm{BHL}\ll\Rp$) explored in similar studies, in which the bow shock envelops the recompression shock \citep{Thun+16,Prust+Bildsten24}.

The middle panel of Fig.~\ref{fig:shock_sweeping} shows the forward shock propagating through the planet. Its curvature initially matches that of the planetary surface. At the same time, a rarefaction wave propagates into the planet from its back side in response to the movement of the background medium away from the planetary surface. This is visible as a pale-red crescent-shaped region in the left panel. The forward shock and rarefaction wave then impinge on one another, starting from the outer region of the planet. This interaction decelerates the forward shock and reverses its curvature, as shown in the middle panel.

The right panel shows the quasi-steady state reached long after the forward shock has swept through the planet. Kelvin--Helmholtz vortex sheets on the planet's surface induce oblique shocks that travel downstream. Unlike the flow past a smooth sphere, no single recompression shock is present at late times. Instead, flow recompression occurs through local oblique shocks that collectively form a compressive region. Partial reflection of the bow shock at the domain boundaries is also visible, impinging on the ablated material far downstream of the planet.

\subsection{Mass ablation}
\label{subsec:ablation}
The vortex sheets also contribute to mass ablation by mixing planetary material into the background wind. As a result, the amount of mass constituting the planet is not uniquely defined. Nonetheless, we devise an iterative procedure that provides a smoothly varying estimate of both the planet's centre of mass, $\mathbf{r}_\mathrm{CM}$, and its total mass across simulation snapshots. The algorithm is fully described in Appendix \ref{app:mass_calculation}.
\begin{figure}
    \includegraphics[width=\columnwidth]{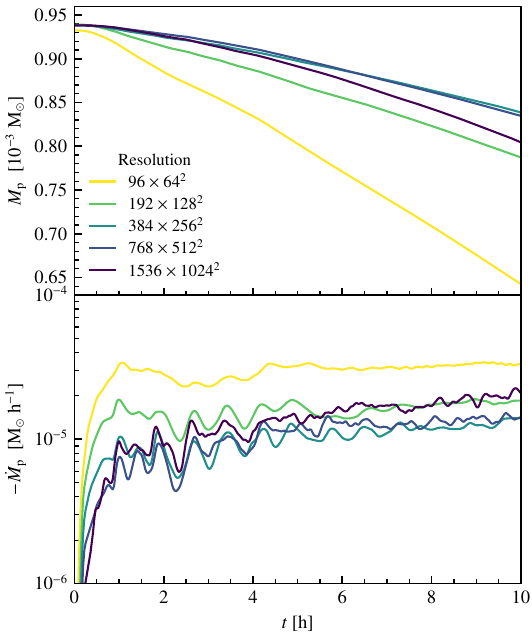}
    \caption{Top panel: Evolution in planet mass for the fiducial wind parameters ($f_\mathrm{m}=f_\mathrm{e}=1$, $\mathcal{M}_\infty=2$) at different grid resolutions, spanning a factor of $2^4$ in linear resolution. Bottom panel: Evolution in mass-ablation rate for the same simulations.}
    \label{fig:planet_mass}
\end{figure}

Figure~\ref{fig:planet_mass} shows the evolution of the planet mass and ablation rate for the fiducial parameters ($f_\mathrm{m}\,{=}\,f_\mathrm{e}\,{=}\,1$, $\machinf\,{=}\,2$) at different resolutions. The ablation rate reaches $\Mpdot\,{\sim}\,10^{-5}\Msunhr$ by $t\,{=}\,1\hr$, which corresponds to slightly more than ten wind-crossing times ($\Rp / v_\infty$). The mass-loss process is not uniform---the lower panel shows variations in \Mpdot due to oscillations excited in the planet, most prominently at early times. On longer timescales, the average \Mpdot increases gradually. This could be caused by the planet's radial expansion by up to ten percent, which, along with mass loss, reduces the escape velocity at the planet's surface. As a result, the simulations can only reach a quasi-steady state defined on timescales that are shorter than the mass-loss timescale, $\Mp/(\Mpdot)$.

For the fiducial wind parameters, \Mpdot exhibits non-monotonic convergence. It initially decreases roughly by a factor of three as the linear resolution increases by a factor of four from the $96\times64^2$ grid to the $384\times256^2$ grid. At higher resolutions, however, the average \Mpdot increases slightly again, especially at late times as seen in the $1536\times1024^2$ simulation. Unless further non-monotonic behaviour appears at even higher grid resolution, we conservatively estimate that the ablation rate is at least ${\sim}\,10^{-5}\Msunhr$ for the fiducial set of wind parameters. In Sect.~\ref{subsec:applications}, we show that this level of mass loss can lead to gradual but complete destruction of an engulfed planet.

In the global planetary engulfment simulations by \cite{Lau+25}, a mass-loss rate of ${\approx}\,5\times 10^{-5}\Msunhr$ is reported when the hot Jupiter experiences the same conditions as our fiducial case (at an orbital separation of 2.8\Rsun). This is in reasonable agreement despite the different setups and numerical methods. However, the mass-loss rate reported by \cite{Lau+25} is not converged, decreasing with increasing resolution. In their highest-resolution smoothed particle hydrodynamics simulation, the hot Jupiter is initially resolved by 30,000 particles, resulting in a comparable discretisation length scale to our $384\times256^2$-grid run, where the planet is initially resolved by 74,392 cells. A controlled, resolution-specific comparison between a finite-volume code and a smoothed particle hydrodynamics code is beyond the scope of this work, but has been performed for selected problems \citep[e.g][]{Price+Federrath10,Tricco19}.

\subsection{Dependence on the wind momentum and energy fluxes}
\label{subsec:ablation_mechanism}

\begin{figure*}
    \centering
    \includegraphics[width=\textwidth]{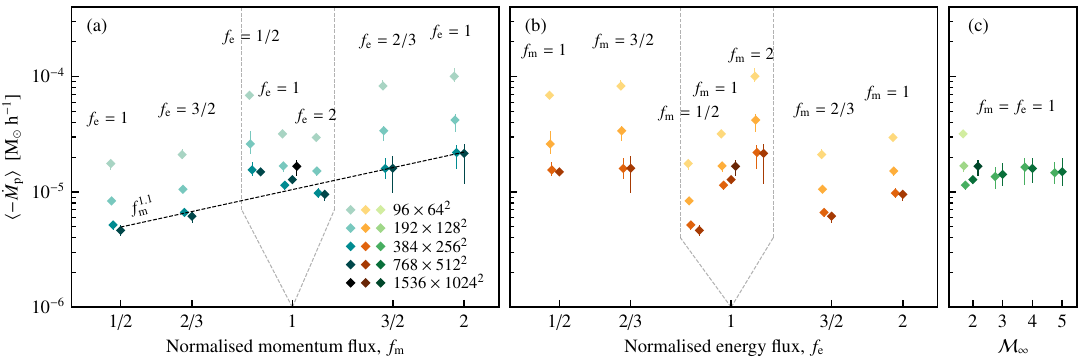}
    \caption{Quasi-steady planet ablation rate, $\langle\Mpdot\rangle$, as a function of $f_\mathrm{m}$ (Panel (a)), $f_\mathrm{e}$ (Panel (b)), and \machinf (Panel (c)) at different grid resolutions. The error bars indicate 1$\sigma$ uncertainties due to temporal fluctuations and any secular growth of \Mpdot (Appendix \ref{app:steady-state}). To prevent overlapping, data markers with the same abscissa are separated in the dashed regions and a small horizontal displacement is also applied elsewhere.
    The dashed black line shows a least-squares fit of $\langle\Mpdot\rangle\propto f_\mathrm{m}^\alpha$ with $\alpha\,{=}\,1.1^{+0.2}_{-0.3}$.}
    \label{fig:ablation_rate}
\end{figure*}

We now examine how the ablation rate varies with the normalised wind momentum flux ($f_\mathrm{m}$) and energy flux ($f_\mathrm{e}$), which together parametrise the wind density and velocity. To facilitate comparison between simulations, we computed quasi-steady values, $\langle\Mpdot\rangle$, along with $1\sigma$ confidence intervals using the method described in Appendix \ref{app:steady-state}. The results are listed in Table~\ref{tab:summary} and shown as a function of $f_\mathrm{m}$ and $f_\mathrm{e}$ in Fig.~\ref{fig:ablation_rate}. The ablation rate generally decreases with increasing grid resolution. In all cases, $\langle\Mpdot\rangle$ has converged or is close to convergence at a grid resolution of $384\times256^2$.

Figure~\ref{fig:ablation_rate}(a) shows that $\langle\Mpdot\rangle$ scales positively with $f_\mathrm{m}$ for simulations with the same upstream Mach number, fixed at the fiducial value of $\machinf\,{=}\,2$. A least-squares fit to the highest-resolution results gives
\begin{align}
    \langle\Mpdot\rangle\,/\,(\mathrm{M}_{\odot}~\mathrm{h}^{-1}) =10^k\,f_\mathrm{m}^\alpha,
\end{align}
with $\alpha\,{=}\,1.08^{+0.19}_{-0.27}$ and $k\,{=}\,-4.981^{+0.031}_{-0.034}$, where the uncertainties reflect $1\sigma$ stochastic variation in \Mpdot. This fit is shown by the dashed black line.

Comparing individual examples further supports this trend. The case with $f_\mathrm{m}\,{=}\,3/2$, $f_\mathrm{e}\,{=}\,2/3$ has a 50\% higher momentum flux and 50\% lower energy flux than the fiducial case ($f_\mathrm{m}\,{=}\,f_\mathrm{e}=1$), and exhibits a 2.6 times higher ablation rate than the opposite case ($f_\mathrm{m}\,{=}\,2/3$, $f_\mathrm{e}\,{=}\,3/2$). Panel~(b) shows that there is indeed no monotonic dependence of $\langle\Mpdot\rangle$ on $f_\mathrm{e}$. The dashed region in Panel~(a) contains simulations with fixed $f_\mathrm{m}$ but different $f_\mathrm{e}$. While $\langle\Mpdot\rangle$ does vary significantly among these cases, it in fact decreases with $f_\mathrm{e}$. We therefore conclude that momentum flux, rather than energy flux, primarily controls the ablation rate.

In simulations with relatively high $\rho_\infty$, we find that the combination of the turbulent wake and overdense post-shock flow leads to gravitational collapse. Pressure forces are insufficient to resist contraction, resulting in the Jeans instability. For the fiducial wind parameters, the Jeans length may be approximated as
\begin{align}
    \lambda_\mathrm{J} = c_{s,\infty}\bigg( \frac{\pi}{G\rho_\infty} \bigg)^{1/2}
    = 10f_\mathrm{m}^{-5/2} f_\mathrm{e}^2 (\mathcal{M}_\infty/2)^{-1}~\mathrm{R}_\odot,
\end{align}
where $c_{s,\infty}$ and $\rho_\infty$ are expressed in terms of $f_\mathrm{m}$, $f_\mathrm{e}$, and \machinf. This confirms that $\lambda_\mathrm{J}$ exceeds the size of the computational domain for the fiducial simulation, so only stable perturbations are possible. However, the scaling also implies $\lambda_\mathrm{J}$ decreases to ${\sim}\,1\Rsun$ for the $f_\mathrm{m}\,{=}\,3/2$, $f_\mathrm{e}\,{=}\,2/3$ and $f_\mathrm{m}\,{=}\,2$, $f_\mathrm{e}\,{=}\,1$ cases. In these simulations, the downstream density grows over time, increasing gravitational drag beyond the simple estimate of Eq.~(\ref{eq:BHL}). This additional drag displaces the planet from the region of highest resolution, as the imposed frame acceleration only compensates for the initial pressure drag. Eventually, the collapse draws inflows from the domain boundaries, and we terminate the simulation before reaching $t\,{=}\,10\hr$. We confirmed that this behaviour is absent when self-gravity is disabled and the planet is maintained in approximate hydrostatic balance using a fixed potential. In these cases, the simulations are instead performed on a uniform Cartesian grid in the inertial frame and show qualitatively similar evolution. We calculated $\langle\Mpdot\rangle$ and other quantities during the quasi-steady phase maintained at earlier times.

\subsection{Efficiency of pressure drag and ablation}
\label{subsec:efficiency}
The wind momentum intercepted by the planet induces both pressure drag and mass ablation, among other processes. To disentangle these different components, we computed the relative contributions of each term in the momentum equation. We carry out a momentum decomposition by integrating the $x$-component of Eq. (\ref{eq:momentum_conservation}) over a spherical volume, $\mathcal{V}$, that encloses the planet. Applying the divergence theorem, we obtain
\begin{multline}
    \underset{(\mathrm{i})}{\frac{\mathrm{d}}{\mathrm{d}t} \int\limits_{\mathcal{V}}\!\rho v_x \mathrm{d}\mathcal{V}}= 
    - \underset{(\mathrm{ii})}{\int\limits_{\partial \mathcal{V}}\!\rho v_x \bm{v}\!\cdot\!\mathrm{d}\bm{S}}
    - \underset{(\mathrm{iii})}{\int\limits_{\partial \mathcal{V}} \!p \hat{\bm{x}}\!\cdot\!\mathrm{d}\bm{S}}
    - \underset{(\mathrm{iv})}{\int\limits_\mathcal{V}\!\rho a_\mathrm{p,x} \mathrm{d}\mathcal{V}}
    - \underset{(\mathrm{v})}{\int\limits_{\mathcal{V}}\!\rho\frac{\partial\phi}{\partial x} \mathrm{d}\mathcal{V}},
    \label{eq:momentum_decomposition}
\end{multline}
The interpretation of each term is given below:
\begin{enumerate}[(i)]
    \item Time derivative of the total $x$-momentum within $\mathcal{V}$.
    \item Net advection rate of $x$-momentum into $\mathcal{V}$.
    \item Force arising from an asymmetric pressure distribution over the sphere $\partial \mathcal{V}$. The dominant contribution comes from the post-shock pressure between the planet's front face and the bow shock, while the pressure in the wake is small in comparison. This term gives rise to pressure or form drag acting on the planet.
    \item Fictitious force associated with the uniformly accelerating frame. The acceleration, $a_{\mathrm{p},x}$, is chosen to hold the planet still, offsetting the acceleration due to pressure drag in the inertial frame.
    \item Gravitational force due to asymmetries in the background density. When restricted to planetary material, the integral gives the gravitational drag acting on the planet.
\end{enumerate}
The interpretation of the drag force terms does not depend on how closely the volume traces the planet's surface, so long as the bow shock is excluded. The sphere was chosen to have a fixed radius of 0.12\Rsun and is positioned along $\hat{\mathbf{x}}$ such that it just contains the planet's front face.

\begin{figure}
    \includegraphics[width=\linewidth]{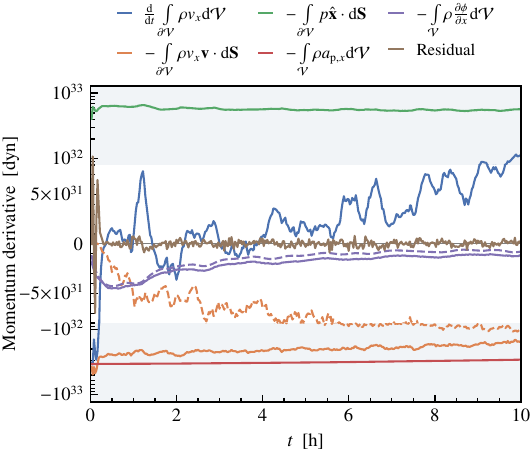}
        \caption{Terms in the integrated momentum equation (Eq.~(\ref{eq:momentum_decomposition})) for the fiducial simulation. Dashed lines show the contributions from planetary material and the brown line shows the residual. To facilitate comparison between terms, the ordinate switches from a linear to a logarithmic scale in the shaded regions.}
    \label{fig:momentum_decomposition}
\end{figure}

Figure~\ref{fig:momentum_decomposition} shows selected terms for the fiducial simulation. We evaluated the volume integrals (terms (i), (iv), and (v)) using a triple Riemann sum and approximated the time derivative in term (i) with first-order backward differencing. The surface integrals, terms (ii) and (iii), were evaluated using 131st-order Lebedev quadrature \citep{Lebedev99}. For terms (ii) and (v), we also explicitly show the contribution from planetary material (dashed lines), obtained by including $\psi$ in the integrand.

The dominant term is term (iii) (green line), which is mainly in balance with the negative contributions from terms (ii) (orange) and (iv) (red) that remove momentum from the control volume. The remaining terms contribute less than ${\sim}\,10\%$ of the dominant term. For example, term (i) (blue) is at most ${\approx}\,17\%$ of term (iii) (green), as expected since the time derivative vanishes in a completely steady state. The residual (brown) is at a level of ${\lesssim}\,1\%$. The dashed orange line represents the momentum carried out of the control volume, $\mathcal{V}$, by ablated planetary material. This term is also small, indicating that momentum loss due to mass ablation is relatively inefficient. The gravitational term (purple), is even smaller, verifying the limited role of gravitational forces in the fiducial case. However, it becomes more significant in simulations with denser winds. Notably, the sign of this term is negative, giving rise to so-called negative dynamical friction that removes $x$-momentum. This acceleration effect has been studied for objects with mass outflows moving relative to a background medium \citep{Shima+86,Inaguchi+86,Gruzinov+20}, as well as in the context of planetary engulfment \citep{Yarza+23}, and is due to the formation of an underdense wake behind the embedded object. Caution is needed in interpreting the sign or magnitude of gravitational drag, as it is potentially sensitive to density perturbations on the largest scale, which is set by the size of the stellar envelope but limited in our simulations by the computational domain size.

We use term (ii) to define a dimensionless coefficient that quantifies the efficiency of mass ablation. To isolate the net $x$-momentum carried away by ablated planetary material, we include the passive scalar, $\psi$, in the integrand. We then normalise the integral by the rate at which $x$-momentum is intercepted by the planet, resulting in the expression
\begin{align}
    \eta_\mathrm{abl} &:= \frac{1}{A_\mathrm{eff}\rho_\infty \Delta v^2} \int\limits_{\partial \mathcal{V}}\rho \psi v_x \mathbf{v}\cdot \mathrm{d}\bm{S},
    \label{eq:eta_ablation}
\end{align}
where $A_\mathrm{eff}$ is the planet's cross-sectional area, computed by integrating $\psi$ over the $yz$-plane containing the planet's centre of mass, i.e., $A_\mathrm{eff}\,{=}\,\iint_{x=x_\mathrm{CM}} \psi \mathrm{d}y\mathrm{d}z$. Its product with the momentum flux, $\rho_\infty \Delta v^2$, gives the rate at which momentum is intercepted by the planet, where $\Delta v$ is the $x$-velocity contrast between the wind and the planet's centre of mass. In the accelerated reference frame, the planet is nearly stationary by construction, and so $\Delta v \approx v_\infty$. However, in the dense-wind cases, the overdense wake behind the planet gravitationally accelerates the incident flow. We therefore computed $\Delta v$ using the velocity difference between the planet's centre of mass and the flow immediately upstream of the bow-shock apex. Because the dense-wind simulations were performed in the inertial frame, the planet velocity cannot be neglected. In these cases, the advection velocity $\mathbf{v}$ in Eq.~(\ref{eq:eta_ablation}) is replaced by the velocity contrast, $\mathbf{v}-v_\mathrm{p,x}\hat{\mathbf{x}}$. This correction arises from term (i) in Eq.~(\ref{eq:momentum_decomposition}) when accounting for the control volume's motion using the Reynolds transport theorem.

Next, we define the pressure-drag coefficient using the planet's acceleration:
\begin{equation}
    \eta_\mathrm{pres} := \frac{1}{A_\mathrm{eff}\rho_\infty \Delta v^2}
    \Bigg[~
        \Mp \Bigg(\frac{\partial^2 x_\mathrm{CM}}{\partial t^2} + a_{\mathrm{p},x}\Bigg) - \int\limits_{\mathcal{V}}-\rho\psi\frac{\partial\phi}{\partial x} \mathrm{d}\mathcal{V}
    ~\Bigg]~.
    \label{eq:eta_ram_pressure}
\end{equation}
The first two terms in parentheses give the planet's centre-of-mass acceleration in the inertial frame (corresponding to terms (i) \& (iv)). The third term subtracts the contribution from gravitational drag (term (v)), which becomes significant in the dense-wind simulations that produce a gravitationally collapsing wake. Studies that model the planet as a rigid sphere \citep[e.g.][]{Yarza+23,Prust+Bildsten24} instead compute the pressure drag directly from the surface integral, term (iii). This approach is not suitable in our case, as the planet's boundary is not well defined. It is common in the literature to include an additional factor of 1/2 in the normalisation \citep[e.g.][]{Loth+21}, but we do not include this to be consistent with the definition of $\eta_\mathrm{abl}$.

\begin{figure*}
    \centering
    \includegraphics[width=\textwidth]{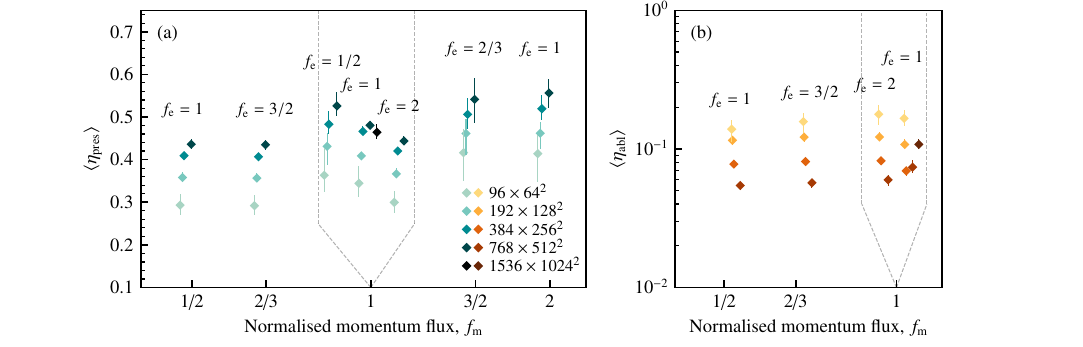}
    \caption{Quasi-steady values of the ablation and pressure-drag coefficients (Eq.~(\ref{eq:eta_ablation}) \& (\ref{eq:eta_ram_pressure})), for different simulations as a function of the normalised momentum flux. The error bars and the dashed region have the same meaning as in Fig.~\ref{fig:ablation_rate}.}
    \label{fig:efficiency}
\end{figure*}

Figure~\ref{fig:efficiency} shows quasi-steady values of $\eta_\mathrm{abl}$ and $\eta_\mathrm{pres}$ across the different simulations, plotted as functions of $f_\mathrm{m}$. We applied the same method for extracting a quasi-steady mass-loss rate as described in Appendix~\ref{app:steady-state}, with an example shown in Fig.~\ref{fig:fitting_example}. We do not include $\langle \eta_\mathrm{abl}\rangle$ for the dense-wind cases because we find that $\eta_\mathrm{abl}$ does not reach a quasi-steady value. This is likely related to the onset of the Jeans instability discussed in Sect.~\ref{subsec:ablation_mechanism}, where the growing overdense wake could enhance mass stripping. For the fiducial case, we find $\langle\eta_\mathrm{pres}\rangle=0.46^{+0.02}_{0.02}$, in good agreement with experiments and hydrodynamical simulations, which determine the drag coefficient for a smooth sphere to be close to 1/2 for transonic and supersonic flows in the subcritical regime, where the Reynolds number ranges from a few thousand to $\sim 2.5\times 10^5$ \citep[e.g.][]{Miller+Bailey79,Loth+21}. 

The three dense-wind cases ($f_\mathrm{m}\,{>}\,f_\mathrm{e}$) have slightly larger values of $\langle\eta_\mathrm{pres}\rangle$, ranging from 0.53 to 0.56, although these measurements carry larger uncertainties due to increased stochastic variability. This modest enhancement may indicate the start of a transition from classical hydrodynamical drag to a regime where self-gravity modifies the flow geometry and pressure distribution compared to that observed in laboratory experiments. In these simulations, it becomes less appropriate to neglect the gravitational acceleration and focussing that the upstream flow would have experienced as it travels from infinity to the inlet boundary. We estimate based on ballistic motion that this effect is minor for the fiducial and low-density wind cases, where the upstream flow velocity increases by $\lesssim 10\%$ from the planet's gravity. However, in the most extreme dense-wind case, this increase reaches $\approx 40\%$ because of the lower wind velocity and the larger Bondi-Hoyle-Lyttleton radius relative to the planet's radius. The flow continues to accelerate within the computational domain before reaching the apex of the bow shock, increasing by a further 10--20\% from the gravity of the overdense wake. It is therefore important to apply the drag coefficients using the true incident momentum flux, evaluated at the bow-shock apex, for the dense-wind cases.

The ablation coefficients tend to be significantly smaller, ranging from 0.054 to 0.11, indicating that only a small fraction of the intercepted momentum is expended in mass ablation. Figure~\ref{fig:efficiency} shows little variation in $\langle \eta_\mathrm{abl}\rangle$ with $f_\mathrm{m}$ and $f_\mathrm{e}$, although a larger set of simulations is required to confirm this. For the fiducial simulation, we obtain $\langle \eta_\mathrm{abl}\rangle\,{=}\,0.108^{+0.010}_{-0.007}$. The non-monotonic convergence in $\langle\Mpdot\rangle$ is also reflected in $\langle \eta_\mathrm{abl}\rangle$: it initially decreases with increasing resolution, but rises again beyond the $384\times256^2$ grid. In Sect.~\ref{subsec:applications}, we nevertheless demonstrate how these efficiency parameters can be applied to model the inspiral of an engulfed planet within a stellar envelope, noting that improved convergence and exploration of a wider parameter region would enable more accurate predictions.

\subsection{Effect of the wind Mach number}
\label{subsec:mach}
To briefly explore the effect of the upstream Mach number, we performed simulations with $\machinf\,{=}\,3,4,5$ while keeping $f_\mathrm{m}\,{=}\,f_\mathrm{e}\,{=}\,1$ (rows (viii)--(x) in Table~\ref{tab:summary}). This is achieved by decreasing the wind pressure while keeping $\rho_\infty$ and $v_\infty$ fixed, thereby decreasing the sound speed. Figure~\ref{fig:ablation_rate}(c) shows that these variations produce no significant change in $\langle\Mpdot\rangle$, with the best-fit values agreeing within 18\%. This variation is much smaller than that induced by changes in $f_\mathrm{m}$, indicating that the ablation rate depends only weakly on \machinf. In contrast, Fig.~\ref{fig:mach-comparison} shows that \machinf influences the flow morphology and shock structure. In line with expectations, the half-opening angle of the Mach cone decreases with \machinf, closely following $\arcsin(\mathcal{M}_\infty^{-1})$. In Appendix \ref{app:standoff}, we show that the shock stand-off distance varies with \machinf in a manner consistent with laboratory experiments and numerical simulations.

\section{Discussion}
\label{sec:discussion}

\subsection{Analytical theory for mass ablation}
\label{subsec:kelvin_helmholtz}
The growth and shedding of vortices observed at the planet's surface suggest mass ablation taking an origin in the Kelvin--Helmholtz instability. Based on this suggestion, we derive an analytical expression for the ablation rate based on the growth rate of the Kelvin--Helmholtz vortices. We show that this adds theoretical support for the scaling of $-\dot{M}_\mathrm{p}$ with the wind momentum flux and allows us to connect the ablation efficiency, $\eta_\mathrm{abl}$, to quantities with specific physical interpretations.

We equate the rate at which momentum is carried away by ablated material with the rate at which momentum is imparted by the wind incident on the growing vortex sheets:
\begin{align}
    \Mpdot v_\mathrm{abl} = 2\pi \Rp d \rho_\infty v_\infty^2,
    \label{eq:ablation1}
\end{align}
where $v_\mathrm{abl}$ is the threshold ablation speed and $2\pi \Rp d$ is the annular interaction cross-section from a vortex sheet protruding a distance $d$ from the planet's surface. We treat $v_\mathrm{abl}$ as an uncertain parameter. It must exceed the planet's escape velocity, $v_\mathrm{esc,\,p}\,{=}\,(2GM_\mathrm{p}/\Rp)^{1/2}$, in order for the ablated material to escape the planet's gravity, but is unlikely to be much larger and should not exceed $v_\infty$. Following \cite{Lau+25}, we set $v_\mathrm{abl}\,{=}\,\xi v_\mathrm{esc,\,p}$ where $1~{\leq}~\xi~{\leq}~v_\infty / v_\mathrm{esc,\,p}$. In our simulations, we calculated $\xi$ by dividing the momentum ablation rate from Eq.~(\ref{eq:eta_ablation}) by $\Mpdot v_\mathrm{esc,p}$. We find that $\xi\,{\approx}\,1.0$--$1.5$ in all cases, meaning there is no strong dependence on the wind properties. This differs from the analysis of \cite{Jia+Spruit18}, who instead set $v_\mathrm{abl}\,{\sim}\,v_\infty$.

However, we follow \cite{Jia+Spruit18} in setting $d$ based on the wavelength of the fastest-growing mode that may be amplified to a non-linear state in the planet-crossing time, $\Rp/v_\infty$, assuming the Kelvin--Helmholtz vortex-sheet geometry \citep{Landau44}. This results in $d = (2\pi \varepsilon/n) s\,\Rp$, where $s\,{=}\,(\rho_\mathrm{surf,p}\rho_\infty)^{1/2}/(\rho_\mathrm{surf,p}+\rho_\infty)$ with $\rho_\mathrm{surf,p}$ being the planet's surface density, $\varepsilon$ is the thickness of non-linear vortex sheets developing on the planet's surface in units of the wavelength, and $n$ is the number of $e$-foldings required for the surface to develop into a vortex, which is a saturated non-linear state. Substituting this expression into Eq.~(\ref{eq:ablation1}), we obtain
\begin{align}
    \Mpdot = \frac{4\pi^2\varepsilon}{n\xi} \frac{\Rp^2}{v_\mathrm{esc,\,p}}s~\rho_\infty v_\infty^2.
    \label{eq:ablation_rate}
\end{align}
This expression recovers the nearly linear scaling of \Mpdot with the momentum flux observed in Fig.~\ref{fig:ablation_rate}, although there is an extra factor, $s$, that approaches $(\rho_\infty/\rho_\mathrm{surf,p})^{1/2}$ in the limit of $\rho_\infty\,{\ll}\,\rho_\mathrm{surf,p}$. If $v_\mathrm{abl}$ were assumed to scale as $v_\infty$, the ablation rate would instead scale primarily as the mass flux, $\rho_\infty v_\infty$.

Comparing Eq.~(\ref{eq:ablation_rate}) with Eq.~(\ref{eq:eta_ablation}) provides an expression for the ablation coefficient:
\begin{align}
    \eta_\mathrm{abl} = \frac{4\pi\varepsilon}{n} s.
    \label{eq:eta_ablation_analytical}
\end{align}
This expression depends on physical parameters characterising the Kelvin--Helmholtz instability and the density contrast between the planet's surface and the background, encapsulated in $s$. Substituting values for our fiducial simulation, assuming $\xi\,{=}\,1.1$, and adopting $\epsilon\,{=}\,0.5$ and $n\,{=}\,10$ following \cite{Jia+Spruit18}, Eq.~(\ref{eq:ablation_rate}) yields $\Mpdot\,{=}\,2.1\times 10^{-5}\Msunhr$, which is only 1.3 times larger than the measured value. For the fiducial simulation, $\rho_\infty/\rho_\mathrm{surf,p}\,{\approx}\,0.02$ initially. We verified that $\rho_\mathrm{surf,p}$ does not change by more than a few times 10\% during the simulation, measuring it at the edge of the velocity shear layer on the planet's lateral side. Eq. (\ref{eq:eta_ablation_analytical}) yields $\eta_\mathrm{abl}\,{=}\,0.089$, which is ${\approx}\,0.8$ times the measured value. The agreement is therefore very good given sizeable uncertainties in the values of $\varepsilon$, $n$, and $\xi$.

Equation (\ref{eq:ablation_rate}) adopts the linear growth rate of the Kelvin--Helmholtz instability in the incompressible limit ($\mathcal{M}_\infty \ll 1$). In transonic and supersonic flows, the growth rate is generally suppressed as the upstream flow does not have sufficient time to adjust to the perturbed surface. However, in our limit of $\rho_\infty/\rho_\mathrm{surf,p}\,{\approx}\,0.02$ and at Mach numbers of a few, compressibility effects are minor and only change the maximum growth rate by a few percent \citep{Landau44,Syrovatskii54}.

\subsection{Applications}
\label{subsec:applications}

\begin{figure*}[h!]
    \centering
    \includegraphics[width=\textwidth]{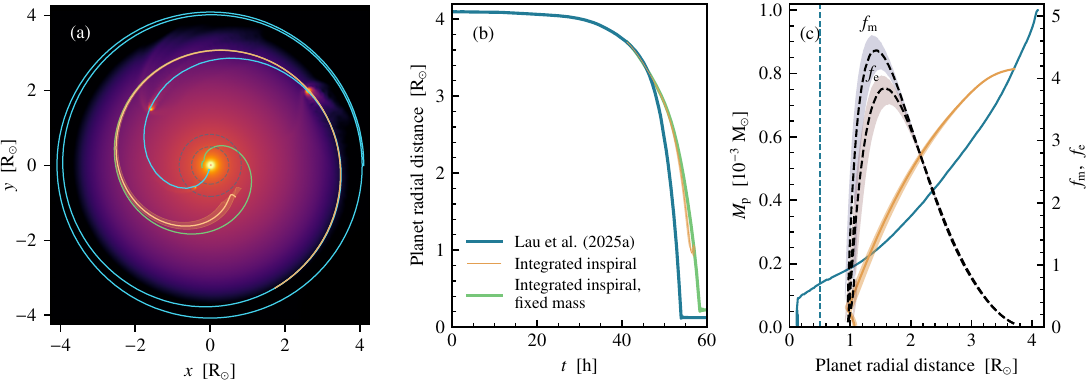}
    \caption{Comparison of the planet's inspiral and mass evolution across several models. The blue line shows results from the 3D simulations by \cite{Lau+25}, the orange line shows the integrated inspiral using the pressure-drag and ablation coefficients derived from the fiducial simulation, and the green line shows an integrated inspiral that neglects mass ablation. Panel (a): Inspiral trajectories plotted over stacked density slices from the 3D simulation. The innermost dashed circle marks the base of the convective envelope, while the other circle indicates the expected depth for global ram-pressure disruption. Panel (b): Time evolution of the planet's radial distance from the centre of the red giant. Panel (c): Evolution of the planet's mass (left axis, solid lines) and the normalised momentum and energy fluxes (right axis, black dashed lines). The blue vertical dashed line shows the base of the convective envelope. The shaded bands around the orange line and the black dashed lines reflect deviations due to the $1\sigma$ uncertainty in the drag and ablation coefficients.}
    \label{fig:1d_3d_traj_comparison}
\end{figure*}

A primary motivation for this study is to enable modelling the observational consequences of planetary engulfment across a wide parameter space using physically motivated prescriptions that are calibrated by self-consistent simulations. A common approach involves integrating a planet's equations of motion within a spherically symmetric stellar model, which is significantly simpler and computationally less expensive than full hydrodynamical simulations. These models neglect the back-reaction of the planet on the stellar structure, which is an adequate assumption in most cases given the planet's insignificant gravitational influence on stellar layers interior to its orbit \citep[e.g.][]{Lau+25}.

However, existing models typically assume that the planet remains intact over much of the inspiral and is only destroyed upon reaching a specific depth, in a catastrophic manner due to ram pressure or tidal disruption \citep[e.g.][but, see \citealt{Church+20}]{Jia+Spruit18,OConnor+23,Yarza+23}. Our simulations indicate that continuous mass ablation may play a significant role before such catastrophic destruction. We therefore propose incorporating gradual mass ablation into the system of differential equations describing the inspiral of an engulfed planet. These equations make use of the mass-ablation and pressure-drag coefficients, $\eta_\mathrm{abl}$ and $\eta_\mathrm{pres}$, that we determined from our wind-tunnel simulations:
\begin{align}
    \dot{\mathbf{r}} &= \mathbf{v}, \label{eq:integrate1} \\
    M_\mathrm{p}\dot{\mathbf{v}} &= -\frac{Gm(r)}{r^2} [M_\mathrm{p}-\rho(r)V] \hat{\mathbf{r}} - \eta_\mathrm{pres} A \rho(r)|\mathbf{v}|\mathbf{v}, \label{eq:integrate2} \\
    \dot{M}_\mathrm{p} &= -\frac{\eta_\mathrm{abl}A\rho|\mathbf{v}|^2}{\xi v_\mathrm{esc,\,p}}, \label{eq:integrate3}
\end{align}
where $\mathbf{r}\,{=}\,(x,y)$ is the planet's position, $r\,{=}\,|\mathbf{r}|$, and $\mathbf{v}$ is its velocity (assuming a non-rotating star). The planet has cross-sectional area $A\,{=}\,\pi \Rp^2$, volume $V\,{=}\,4\pi \Rp^3/3$, and surface escape velocity $v_\mathrm{esc,\,p}\,{=}\,(2G\Mp/\Rp)^{1/2}$, computed from the instantaneous planet mass. The stellar structure is described by the enclosed mass $m(r)$ and density $\rho(r)$. The terms on the right-hand side of Eq.~(\ref{eq:integrate2}) correspond to gravitational acceleration (including buoyancy) and pressure drag.

We applied these equations to model the planetary engulfment scenario simulated by \cite{Lau+25}. The stellar structure was taken from their $1\Msun$ red giant model evolved to a radius of 4\Rsun. The planet's initial position, velocity, and mass were set to the values in the simulation at a separation of $3.7\Rsun$ from the centre of the red giant. For simplicity, we fixed \Rp to its initial value when evaluating $A$, $v_\mathrm{esc,\,p}$, and $V$, although this likely leads to an overestimate of drag and ablation at late times. For an illustrative calculation, we adopted fixed values of $\eta_\mathrm{pres}\,{=}\,0.464$, $\eta_\mathrm{abl}\,{=}\,0.108$, and $\xi\,{=}\,1.2$ from the fiducial simulation. An accurate inspiral model would require extensive parameter calibration across a sufficiently large set of simulations, which is beyond the scope of this work.

We integrated Eq.~(\ref{eq:integrate1})--(\ref{eq:integrate3}) using the fourth-order Runge--Kutta--Fehlberg (RKF45) method \citep{Fehlberg69} with an adaptive time step. Figure~\ref{fig:1d_3d_traj_comparison} compares the planet's trajectory in the $xy$-plane to that obtained by \cite{Lau+25}. The shaded bands indicate deviations due to $1\sigma$ uncertainties in $\langle\eta_\mathrm{abl}\rangle$ and $\langle\eta_\mathrm{pres}\rangle$. The trajectory terminates at a radius of 1.1\Rsun, once the planet's mass has been fully ablated. This occurs at the end of a hook-shaped segment, where the planet rises buoyantly after overshooting its neutral-buoyancy depth. In reality, the effect of buoyancy is reduced because the planet's radius is also expected to decrease from mass loss and compression. We also computed a model that neglects mass ablation (green line). It follows a similar inspiral trajectory, but reaches deeper layers beneath the convective envelope. This is because the planet retains its initial mean density, requiring greater depths to achieve neutral buoyancy.

We now compare the integrated inspiral with the results of \cite{Lau+25}, noting that their hydrodynamical simulations are not fully converged. Panel~(b) shows that the orbital separation evolves in reasonable agreement, although the integrated model produces a shallower inspiral. \cite{Lau+25} showed that increasing the resolution reduces the inspiral rate, which would bring the two results into closer agreement. As seen from Panel~(c), both models predict complete or nearly complete planet ablation within the convective envelope (outside the inner dashed circle in Panel~(a)). However, \cite{Lau+25} find higher ablation rates at early times and lower rates at late times compared to the integrated inspiral. As discussed by \cite{Lau+25}, the ablation rate is likely overestimated at early times. At late times, the 1D integration likely overestimates the ablation rate by neglecting the reduction in the planet radius's cross-section as a function of mass and depth \citep[e.g.][]{Jia+Spruit18}.

We can also inspect how $f_\mathrm{m}$ and $f_\mathrm{e}$ vary during the inspiral to retrospectively check whether our wind-tunnel simulations cover an appropriate parameter range. Figure~\ref{fig:1d_3d_traj_comparison}(c) shows that $f_\mathrm{m}$ and $f_\mathrm{e}$ (dashed lines) evolve closely together for most of the inspiral. They start close to zero at the stellar surface where $\rho$ is small, increase to a maximum at $1.4$--$1.6\Rsun$, where they differ by up to 10--20\%, and then decline as $|\mathbf{v}|$ reaches zero. Thus, for our specific red giant and hot Jupiter models, the simulations cover the relative deviations between $f_\mathrm{m}$ and $f_\mathrm{e}$, spanning $f_\mathrm{m}/f_\mathrm{e}\,{=}\,1/2$--2. However, their absolute values exceed the explored range, with $f_\mathrm{m}$ and $f_\mathrm{e}$ ranging from 0.05 to 3.6 and 4.1 respectively. In the last 0.5\Rsun of the inspiral, with 30\% of its mass remaining, the planet's inspiral also becomes subsonic. This highlights the need to systematically map the variation of $\eta_\mathrm{abl}$ and $\eta_\mathrm{pres}$ over a broader region of $(f_\mathrm{m},f_\mathrm{e},\machinf)$ parameter space, including an investigation of ablation by subsonic winds.

In light of these uncertainties, we conservatively estimate that a significant fraction of the planet's mass is deposited within the convective envelope. For complete dissolution of a Jupiter-mass planet with meteoritic lithium abundance, the surface lithium abundance is estimated to increase by $\Delta A(\mathrm{Li})\,{=}\,0.14\dex$, following the calculation by \cite{Lau+25}. The lifetime of this enrichment signature is set by the $^7\mathrm{Li}(\mathrm{p},\alpha)\alpha$ burning timescale. Adopting the expression from \cite{Andrassy+Spruit13} and the density, temperature, and hydrogen mass fraction at the base of the convective envelope, we obtain a burning timescale of ${\sim}\,10^6\yr$. Because burning is confined to a thin layer at the base of the convective envelope, the depletion timescale is longer by roughly the ratio of the convective envelope depth to the scale height of the burning layer, which is a factor of ${\sim}\,10^2$. This yields a depletion timescale of ${\sim}\,10^8\yr$, which is an upper estimate because convective flows are slower near convective boundaries than in the bulk of the envelope \citep[e.g.][]{Viallet+13}. Although this is short compared to the duration of red giant evolution, significantly longer timescales, up to ${\sim}\,10^9\yr$, are expected for less evolved stars \citep{Soares-Furtado+21}.

The gradual mass ablation identified in our simulations could substantially expand the parameter space in which planetary engulfment produces chemically peculiar stars. Conventional models typically assume that planetary material is deposited at a characteristic depth determined by thermal dissolution, ram-pressure disruption, or tidal disruption. In contrast, continuous ablation allows planetary material to be deposited throughout the inspiral, including within the convective zone. For example, Fig.~\ref{fig:1d_3d_traj_comparison}(a) shows the radius at which ram pressure overcomes the planet's binding energy \citep{Jia+Spruit18}, which in our example lies within the convective region. However, for stars where this radius occurs below the convective zone, catastrophic disruption would not produce observable surface enrichment. This effect is expected to be particularly important for less-evolved stars near the main-sequence turnoff and on the subgiant branch, because their shallow convective envelopes reduce dilution of deposited material, while the lower temperatures at the base of their envelopes slow lithium depletion \citep{Soares-Furtado+21}. Moreover, these stars have denser envelopes and higher inspiral velocities, increasing the incident momentum flux on the planet and potentially leading to even higher ablation rates than those measured in our simulations.

\section{Conclusions}
\label{sec:conclusions}
We performed three-dimensional hydrodynamical simulations of a Jupiter-like planet engulfed within a stellar envelope. Using a wind-tunnel setup, we inject stellar gas from one side of the computational domain and model the local flow around the planet with up to $1536\times1024^2$ cells. The planet is held approximately stationary in a uniformly accelerating frame, except in cases where significant, time-dependent gravitational drag develops. Unlike previous studies that represented an embedded stellar object as a point mass or rigid sphere, our simulations fully resolve the hydrodynamic structure of a gas giant planet, enabling self-consistent modelling of mass ablation.

We parametrise the incident flow using its momentum flux ($f_\mathrm{m}$), energy flux ($f_\mathrm{e}$), and Mach number (\machinf). To systematically study planet ablation under different conditions, we performed a series of simulations with different wind properties and analysed the pressure drag, ablation rate, and flow morphology achieved in a quasi-steady state. We list our main conclusions below:
 
\begin{enumerate}[(i)]
    \item A continuous mass-ablation process operates during planetary engulfment. Across the explored parameter space, the quasi-steady mass-ablation rate is $\Mpdot\,{\sim}\,10^{-5}\Msunhr$, and is converged or close to convergence in our simulations.
    \item The momentum flux ($f_\mathrm{m}$) of the incident flow is the primary parameter controlling the mass-ablation rate, with a nearly linear scaling of $\Mpdot\propto f_\mathrm{m}^{1.1}$. At fixed momentum and energy fluxes, there is no significant dependence on the Mach number, \machinf.
    \item The pressure-drag and ablation coefficients, $\eta_\mathrm{pres}$ and $\eta_\mathrm{abl}$, quantify the efficiency with which intercepted momentum produces drag and mass ablation, respectively. In our simulations, $\eta_\mathrm{pres}$ lies in the range 0.44--0.56, in reasonable agreement with laboratory experiments and theoretical models, whereas the ablation efficiency is substantially smaller, spanning 0.054--0.11.
    \item Mass ablation is consistent with being driven by the Kelvin--Helmholtz instability at the interface between the planet and background stellar gas. An analytical model based on the Kelvin--Helmholtz instability growth rate, similar to that of \cite{Jia+Spruit18}, reproduces the mass-ablation rate and ablation coefficient in our fiducial simulation to within a few tens of percent, despite depending on several uncertain physical parameters.
    \item The coefficients $\eta_\mathrm{pres}$ and $\eta_\mathrm{abl}$ can be incorporated into the equations of motion for the inspiral of a mass-losing gas-giant planet within a stellar envelope. In our example, a significant fraction of the planet's mass is ablated within the convective envelope, leading to lithium enrichment at the stellar surface of $\Delta A(\mathrm{Li})=0.1\dex$ that persists for $\sim 10^8\yr$.
    \item The gradual mass-ablation process expands the parameter space in which planetary engulfment can produce chemical enrichment of convective giant envelopes. This effect is especially important for stars near the main-sequence turnoff and early subgiant phase, where conventional disruption criteria based on ram-pressure and tidal forces predict deposition of planetary material deeper than the convective envelope.
\end{enumerate}

While this study considers a limited set of parameter variations aimed at elucidating the mass-ablation mechanism, denser parameter sampling is required to derive accurate functional fits for $\eta_\mathrm{abl}$ and $\eta_\mathrm{pres}$. The application in Sect.~\ref{subsec:applications} suggests that $f_\mathrm{m}$ and $f_\mathrm{e}$ span a wider range of values than we have explored, and \machinf is also expected to be significantly higher during the early stages of the inspiral \citep{Lau+25}. In addition, the simulations may be improved by using a more realistic structure for the Jupiter-like planet and a hydrogen-helium equation of state that captures molecular dissociation and ionisation.

Our framework could be extended to other engulfed objects like brown dwarfs and terrestrial planets, which are also promising candidates for producing chemically peculiar and rapidly rotating stars. Another improvement is to incorporate the density stratification in the wind and rotational effects associated with orbital motion, both of which may be important in certain regimes \citep[e.g.][]{MacLeod&Ramirez-Ruiz15,Yarza+23,Gagnier+26}. We have also identified the potential importance of a Jeans-like collapse behind the bow shock for relatively dense winds. This process has not been reported in previous wind-tunnel studies that neglect self-gravity, and its contribution to gravitational drag and other dynamical effects warrant further investigation. Finally, the assumption of quasi-steady flow may break down during rapid phases of inspiral, motivating simulations with time-dependent upstream conditions.

\begin{acknowledgements}
  We thank the anonymous referee for comments that have improved the manuscript. We also thank R\"{u}diger Pakmor, Sebastian Ohlmann, Ryosuke Hirai, and members of the PSO Group at HITS for useful discussions. M. Y. M. L., R. A., D. G., and F. K. R. acknowledge funding by the European Union (ERC, ExCEED, project number 101096243). Views and opinions expressed are however those of the authors only and do not necessarily reflect those of the European Union or the European Research Council Executive Agency. Neither the European Union nor the granting authority can be held responsible for them. We acknowledge support by the Klaus Tschira Foundation. I. M. acknowledges support from the Australian Research Council (ARC) Centre of Excellence for Gravitational Wave Discovery (OzGrav), through project number CE230100016. The authors acknowledge the Gauss Centre for Supercomputing e.V. (GCS) for providing computing time on the HAWK Supercomputer at High-Performance Computing Center Stuttgart.
\end{acknowledgements}

\bibliography{aa60892-26.bib}
\bibliographystyle{aa}

\begin{appendix}

\section{Grid stretching functions}
\label{app:stretched_grid}
The computational coordinates are mapped from a logically Cartesian space to physical space using normalised power-law stretching functions in each direction. The $x$-nodes are given by
\begin{align}
    x_i &= x_\mathrm{min} + \frac{x_\mathrm{max}-x_\mathrm{min}}{a+b f}
    \Big(x_\mathrm{offset} + a\tilde{x}_i + b\tilde{x}_i^5 \Big),  \hspace{0.5cm} i=0,\dots,N_x,
\end{align}
with the normalised coordinates
\begin{align}
    \tilde{x}_i = \frac{i}{N_x} - \tilde{x}_0
\end{align}
and
\begin{align}
    a &= \frac{f}{1+f}, \\
    b &= 1-a, \\
    x_\mathrm{offset} &= a \tilde{x}_0 + b \tilde{x}_0^5, \\
    f &= (1 - \tilde{x}_0)^5 + \tilde{x}_0^5,
\end{align}
where $x_\mathrm{min}\,{=}-0.4\Rsun$ and $x_\mathrm{max}\,{=}\,2.0\Rsun$ are the lower and upper $x$-bounds, $N_x$ is the number of cells along the $x$-direction, and $\tilde{x}_0\,{=}\,0.35$ is the fractional position about which the stretching is applied, approximately corresponding to the planet's centre of mass. The stretching in the $y$- and $z$-directions is similar, but yields a simpler functional form because $y_\mathrm{max}\,{=}-y_\mathrm{min}$, $z_\mathrm{max}\,{=}\,-z_\mathrm{min}$, and the centre for the stretching is $\tilde{y}_0\,{=}\,\tilde{z}_0\,{=}\,0$:
\begin{align}
    \label{eq:y1}
    y_j &= \frac{y_\mathrm{max}}{2}\Big(\tilde{y}_j + \tilde{y}_j^5\Big),\hspace{0.5cm} j=0,...,N_y
\end{align} 
with the normalised coordinates
\begin{align}
    \label{eq:y2}
    \tilde{y}_j &= -1 + \frac{2j}{N_y}.
\end{align}
The $z$-nodes are constructed identically, and the stretching functions are obtained by replacing $y$ with $z$ in Eq.~(\ref{eq:y1})--(\ref{eq:y2}).

\section{Impact of computational domain size}
\label{app:domain_size}
\begin{figure}
    \centering
    \includegraphics[width=0.95\linewidth]{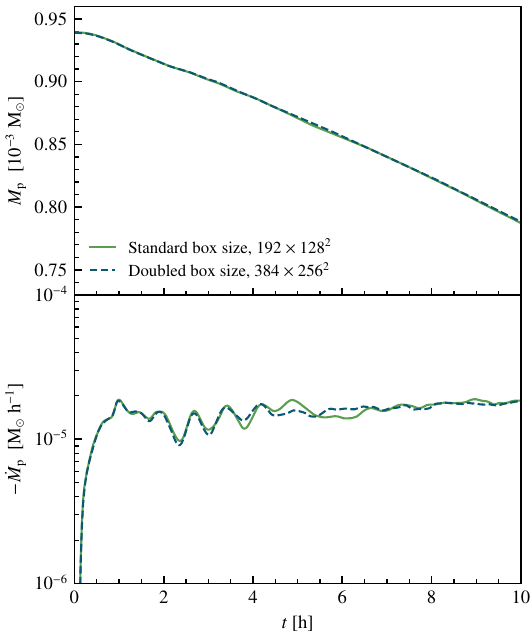}
    \caption{Comparison of mass ablation between a $192\times128^2$-grid simulation (green solid line) and a simulation with twice the box size but the same spatial resolution (blue dashed line). Top: Evolution of planet mass. Bottom: Evolution of mass ablation rate.}
    \label{fig:larger_box_comparison}
\end{figure}

To investigate whether our results are sensitive to the size of the computational domain, we make a comparison to a simulation with twice the domain size. For the larger-box simulation, we used the fiducial wind parameters and a grid resolution of $384\times 256^2$, which has the same spatial resolution as the $192\times 128^2$ grid for the standard domain size. Figure~\ref{fig:larger_box_comparison} shows that there are negligible differences in the evolution of \Mpdot. The pressure drag coefficient obtained with the standard box size is $\langle\eta_\mathrm{pres}\rangle = 0.409^{+0.008}_{-0.006}$, while it is $0.403^{+0.010}_{-0.005}$ with the doubled box size. For the ablation coefficient, the standard box size gives $\langle\eta_\mathrm{abl}\rangle = 0.108^{+0.010}_{-0.006}$, while it is $0.108^{+0.005}_{-0.005}$ with the doubled box size. In either case, there is no statistically significant difference between the best values.

\section{Solving Poisson's equation}
\label{sec:gravity-solver}
We discretise Poisson's equation according to the following finite-difference formula for non-uniform Cartesian grids,
\begin{align}
\label{eq:poisson-FD}
    a_{1,x}\phi_{i-1,j,k} + a_{2,x}\phi_{i,j,k} + a_{3,x}\phi_{i+1,j,k}  \ + &\nonumber \\
     a_{1,y}\phi_{i,j-1,k}  + a_{2,y}\phi_{i,j,k} + a_{3,y}\phi_{i,j+1,k} \  + & \\  
    a_{1,z}\phi_{i,j,k-1} + a_{2,z}\phi_{i,j,k} + a_{3,z}\phi_{i,j,k+1} \    = & \ 4\pi G(\rho_{i,j,k}-\rho_\infty), \nonumber
\end{align}
where
\begin{align}
    a_{1,x} =\ & \frac{2}{\Delta x_-(\Delta x_- + \Delta x_+)}, \\
    a_{2,x} =\ & -\frac{2}{\Delta x_- \Delta x_+}, \\
    a_{3,x} =\ & \frac{2}{\Delta x_+(\Delta x_- + \Delta x_+)}, \\
    a_{1,y} =\ &\frac{2}{\Delta y_-(\Delta y_- + \Delta y_+)}, \\
    a_{2,y} =\ & -\frac{2}{\Delta y_- \Delta y_+}, \\
    a_{3,y} =\ & \frac{2}{\Delta y_+(\Delta y_- + \Delta y_+)}, \\
    a_{1,z} =\ & \frac{2}{\Delta z_-(\Delta z_- + \Delta z_+)}, \\
    a_{2,z} =\ & -\frac{2}{\Delta z_- \Delta z_+}, \\
    a_{3,z} =\ & \frac{2}{\Delta z_+(\Delta z_- + \Delta z_+)}, 
\end{align}
and 
\begin{align}
    \Delta x_- =\ & x_{i,j,k}-x_{i-1,j,k}, \\
    \Delta x_+ =\ & x_{i+1,j,k}-x_{i,j,k}, \\
    \Delta y_- =\ & y_{i,j,k}-y_{i,j-1,k}, \\
    \Delta y_+ =\ & y_{i,j+1,k}-y_{i,j,k}, \\
    \Delta z_- =\ & z_{i,j,k}-z_{i,j,k-1}, \\
    \Delta z_+ =\ & z_{i,j,k+1}-z_{i,j,k}.
    \label{eq:poisson_last}
\end{align}
Here, sets of indices such as $\{i,j,k\}$ denote a cell of the computational grid and $x$, $y$, and $z$ are cell-centred coordinates.

At each time step, we impose boundary conditions for the gravitational potential by computing the monopole term of the multipole expansion of the planet's mass distribution,
\begin{equation}
    \phi(\mathbf{r}_\mathrm{b}) 
    = -\frac{G}{|\mathbf{r}_\mathrm{b}-\mathbf{r}_\mathrm{CM}|}
    \int_{|\mathbf{r}'-\mathbf{r}_\mathrm{CM}|<1.2 R_\mathrm{p}} 
    \big[ \rho(\mathbf{r}')-\rho_\infty \big]\,
    \mathrm{d}^3\mathbf{r}',
\end{equation}
where $\mathbf{r}_\mathrm{b}$ is the boundary location and $\mathbf{r}_\mathrm{CM}$ is the centre of mass of the planet,
\begin{equation}
    \mathbf{r}_\mathrm{CM}=
    \frac{\int_{|\mathbf{r}'-\mathbf{r}_0|<1.2\Rp} 
    \big[ \rho(\mathbf{r}')-\rho_\infty \big]\,
    \mathbf{r}' \,\mathrm{d}^3\mathbf{r}'}
    {\int_{|\mathbf{r}'-\mathbf{r}_0|<1.2\Rp} 
    \big[ \rho(\mathbf{r}')-\rho_\infty \big]\,
    \mathrm{d}^3\mathbf{r}'}.
\end{equation}
Including quadrupole and octupole terms produces no significant differences in the results.

The linear system resulting from the discretisation of Poisson's equation is solved iteratively using a matrix-free implementation of the BiCGSTAB method of \cite{vandervorst1992}. In all simulations, the iteration is terminated when the root-mean-square relative residual,
\begin{equation}
    \label{eq:poisson-error}
    \epsilon = \sqrt{\frac{1}{N_xN_yN_z}
    \sum_{i,j,k}
    \Bigg[
    \frac{(A\phi)_{i,j,k} - 4\pi G (\rho_{i,j,k}-\rho_\infty)}
    {4\pi G\,\mathrm{max}(|\rho_{i,j,k}-\rho_\infty|,\rho_\infty)}
    \Bigg]^2},
\end{equation}
falls below the prescribed tolerance of $10^{-4}$. Here, $(A\phi)_{i,j,k}$ denotes the finite-difference approximation to $\nabla^2\phi$ given by Eq.~(\ref{eq:poisson-FD})--(\ref{eq:poisson_last}). We have verified that the results presented in Sect.~\ref{sec:results} are nearly identical when the tolerance is varied between $10^{-3}$ and $10^{-6}$.

Once Poisson's equation has been solved for $\phi$, the gravitational acceleration $\bm{g}=(g_x,g_y,g_z)=-\nabla\phi$ is approximated using
\begin{align}
    g_x &= -b_x(\phi_{i+1,j,k} - \phi_{i-1,j,k}), \label{eq:grav_x} \\ 
    g_y &= -b_y(\phi_{i,j+1,k} - \phi_{i,j-1,k}), \label{eq:grav_y} \\
    g_z &= -b_z(\phi_{i,j,k+1} - \phi_{i,j,k-1}),\label{eq:grav_z}
\end{align}
with
\begin{align}
     b_x  = \ & \frac{1}{\Delta x_-+\Delta x_+},  \\
     b_y  = \  & \frac{1}{\Delta y_-+\Delta y_+}, \\ b_z  =\  & \frac{1}{\Delta z_-+\Delta z_+}.
\end{align}
This corresponds to a centred finite-difference discretisation of the gradient operator on a non-uniform grid.

The truncation error associated with Eq.~(\ref{eq:poisson-FD}) and Eqs.~(\ref{eq:grav_x})–-(\ref{eq:grav_z}) scales as
\begin{equation}
    \mathcal{O}\!\left(
    \frac{|\Delta^2 \xi_+ - \Delta^2 \xi_-|}
    {\Delta \xi_+ + \Delta \xi_-}
    \right), 
    \qquad \xi = x,y,z.
\end{equation}
Although the scheme is formally first-order accurate on arbitrary non-uniform grids, the first-order truncation term is proportional to differences between the squares of adjacent cell spacings. On a smoothly varying mesh, such as that used in our simulations (Fig.~\ref{fig:grid}), these differences are small, rendering the first-order contribution subdominant. Consequently, the numerical scheme exhibits approximately second-order convergence in practice.

To assess the convergence rate of the gravity solver, we perform a verification test using our wind-tunnel setup on grids of $96\times64^2$, $192\times128^2$, $384\times256^2$, $768\times512^2$, and $1536\times1024^2$ cells. Poisson's equation is solved for the density field at $t=0$ with ${\rho_\infty=\rho_{\infty,0}}$. We adopted monopole boundary conditions and a BiCGSTAB tolerance of $10^{-4}$.

We compute $L_2$ error norms between adjacent resolutions after restricting the finer-grid solution to the coarser grid by averaging neighbouring blocks of ${2 \times 2\times 2}$ cells, 
\begin{equation}
    L_2 = \left\| \phi(N_x,N_y,N_z) - \phi(N_x/2,N_y/2,N_z/2) \right\|_2.
\end{equation}
The error is then normalised by the standard deviation of $\phi$ on the finest grid. Relative errors are computed analogously for the radial component of the gravitational acceleration, $g_r$. 

\begin{figure}
    \centering
    \includegraphics[width=\linewidth]{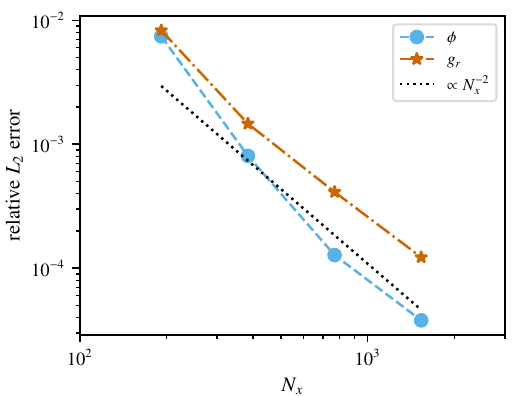}
    \caption{Relative $L_2$ errors of the gravitational potential, $\phi$, and radial gravitational acceleration, $g_r$, as functions of the number of cells $N_x$ along the $x$-direction for our wind-tunnel setup at $t=0$. The black dotted line indicates second-order scaling.}
    \label{fig:L2}
\end{figure}

The results are shown in Fig.~\ref{fig:L2}. The gravitational potential exhibits an apparent convergence rate higher than second-order (${\approx 2.5}$) on the coarser grids, which we attribute to partial cancellation of leading truncation error terms on the smoothly varying mesh. On the finest grids, however, the measured rate decreases to approximately 1.8. This slight degradation is likely related to the fixed BiCGSTAB tolerance of $10^{-4}$, as the algebraic solver error becomes comparable to the discretisation error at high resolution, thereby limiting the observed convergence rate.

\section{Calculation of planet mass}
\label{app:mass_calculation}
The planet's mass, \Mp, and centre of mass, $\mathbf{r}_\mathrm{CM}$, are calculated using an iterative procedure. First, we take the location of the global density maximum as an initial guess for $\mathbf{r}_\mathrm{CM}$. The planet mass is computed as a weighted sum over the product of the cell mass and passive scalar, $\psi$. The weight function is
\begin{align}
    w(r) =
    \begin{cases}
      1, & r \leq R_1, \\
      \frac{r-R_1}{R_2-R_1}, & R_1 < r \leq R_2, \\
      0, & r > R_2, \\ 
    \end{cases}
\end{align}
where $r\,{=}\,|\mathbf{r} - \mathbf{r}_\mathrm{CM}|$ is the radial distance from the planet centre of mass, $R_1\,{=}\,1.2\Rp$, and $R_2{=}\,1.6\Rp$. Thus, full weight is given for $r\,{<}\,R_1$, but the weight decreases linearly between $R_1$ and $R_2$, as opposed to having a sharp mask, which leads to highly stochastic mass measurement. A new centre of mass may be computed with the same weight function, and this procedure is iterated until all components of $\mathbf{r}_\mathrm{CM}$ are converged to within $10^{-3}$. 

\section{Fitting of steady-state quantities}
\label{app:steady-state}
We describe the method used to extract quasi-steady quantities and their uncertainties from each simulation. For the ablation rate, \Mpdot, the planet mass at each time step is first calculated according to the description in Appendix~\ref{app:mass_calculation}. The time derivative is then computed using second-order central finite differencing. To reduce stochasticity in the data, we compute a moving average with a window length of $0.75\hr$. Minor variations in this window length have negligible impact on the best values of $\langle\Mpdot\rangle$ and their $1\sigma$ intervals.

We fit a \ac{KDE} to the data using the \cite{Silverman86} rule of thumb for bandwidth selection. The purpose of this procedure is to further reduce any bias due to noise. We find that \Mpdot reaches a quasi-steady state after $t\,{\approx}\,4\hr$, and construct the \ac{KDE} from data in that window. We adopted the distribution median as the quasi-steady value, $\langle\Mpdot\rangle$, and choose lower and upper limits based on the central $1\sigma$ confidence interval containing equal probability mass above and below the median. 

\begin{figure}
    \centering
    \includegraphics[width=0.95\linewidth]{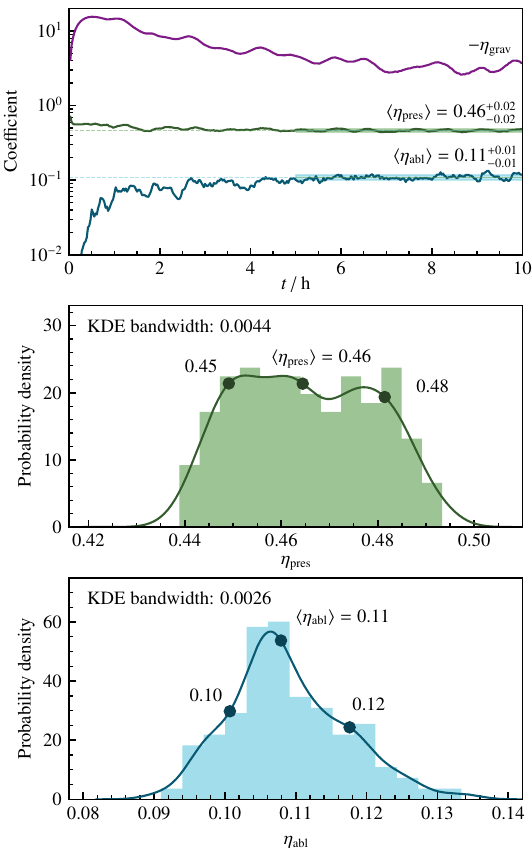}
    \caption{Example fit of the quasi-steady momentum coefficients for the fiducial simulation. Top: Evolution of the gravitational-drag coefficient (purple, Eq.~(\ref{eq:eta_grav_drag})), pressure-drag coefficient (green, Eq.~(\ref{eq:eta_ram_pressure})), and ablation coefficient (blue, Eq.~(\ref{eq:eta_ablation})). The dashed lines show best fits and the shaded bands show $1\sigma$ confidence intervals. The negative of $\eta_\mathrm{grav}$ is plotted. Middle: Histogram of the $\eta_\mathrm{pres}$ time series and its KDE. The markers show the median and 1$\sigma$ confidence interval. Bottom: Histogram and KDE for $\eta_\mathrm{abl}$.}
    \label{fig:fitting_example}
\end{figure}

The same method is applied to determine the steady-state momentum coefficients, $\langle \eta_\mathrm{abl}\rangle$ and $\langle\eta_\mathrm{pres}\rangle$, as well as the shock stand-off distance, $\langle \Delta x_\mathrm{SO}\rangle$. However, unlike for $\langle\Mpdot\rangle$, we do not compute a moving-average of the data as computing these coefficients does not require finite differencing across time steps, and so is not susceptible to the noise it produces. An example of the \ac{KDE} construction is shown in Fig.~\ref{fig:fitting_example}, which shows the time series of $\eta_\mathrm{pres}$ and $\eta_\mathrm{abl}$ in the top panel, followed by their \acp{KDE} in the middle and bottom panels. In general, it is necessary to select a time window for each simulation over which $\eta_\mathrm{pres}$ and $\eta_\mathrm{abl}$ remain approximately steady. In this example, $\eta_\mathrm{pres}$ remains steady throughout the entire simulation, whereas $\eta_\mathrm{abl}$ grows significantly over the first 5 hours. We therefore restrict the data used for the \ac{KDE} construction to the window between $t=5$ and $10\hr$.

Figure~\ref{fig:fitting_example} also includes a gravitational drag coefficient based on term (v) of Eq.~(\ref{eq:momentum_decomposition}):
\begin{align}
    \eta_\mathrm{grav} &:= \frac{1}{\pi R_\mathrm{BHL}^2\rho_\infty \Delta v^2}\int\limits_{\mathcal{V}}-\rho\psi\frac{\partial\phi}{\partial x} \mathrm{d}\mathcal{V},
    \label{eq:eta_grav_drag}
\end{align}
where $A_\mathrm{eff}$ is replaced by $\pi R_\mathrm{BHL}^2$, the area spanned by the Bondi--Hoyle--Lyttleton radius (Eq.~(\ref{eq:BHL})). Figure~\ref{fig:fitting_example} shows that $\eta_\mathrm{grav}$ is of the order of $10^1$ but consistently negative, giving rise to so-called negative dynamical friction \citep[e.g.][]{Gruzinov+20}. However, the magnitude and sign of the drag force may differ when a larger computational domain is used. In most of our simulations, the gravitational drag is only initially negative and transitions to a genuine drag force after a few hours. The parameter $\eta_\mathrm{grav}$ generally exhibits significant time variation and has no steady value.

\section{Dependence of shock stand-off distance on wind properties}
\label{app:standoff}
\begin{figure}
    \includegraphics[width=\linewidth]{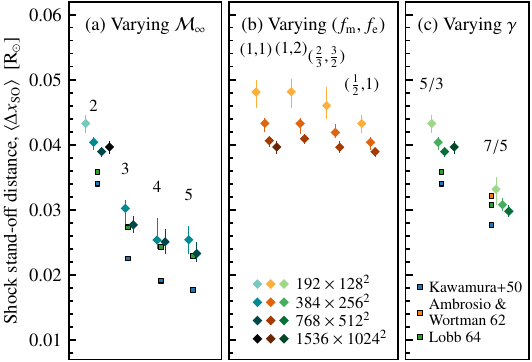}
    \caption{Variation in the quasi-steady shock stand-off distance, $\langle\Delta x_\mathrm{SO}\rangle$, with different properties of the injected wind and grid resolution. Panel (a): Variation in the Mach number, \machinf, with $f_\mathrm{m}\,{=}\,f_\mathrm{e}\,{=}\,1$ fixed. Panel (b): Variation in $(f_\mathrm{m},f_\mathrm{e})$ at fixed $\machinf\,{=}\,2$. Panel (c): Variation in the adiabatic index, $\gamma$, with $(f_\mathrm{m},f_\mathrm{e},\machinf)$ fixed. Square markers show predictions based on experimental data or analytical models (Sect.~\ref{subsec:mach}). Labels above each set of simulations indicate the value of the varied parameter, and the abscissa is arbitrary. As in Fig.~\ref{fig:ablation_rate}, the error bars indicate 1$\sigma$ uncertainties.}
    \label{fig:standoff}
\end{figure}

Laboratory experiments and numerical simulations show that \machinf is inversely correlated with the shock stand-off distance, $\Delta x_\mathrm{SO}$, defined here as the normal distance between the bow shock and the stagnation point on the planet's front face, where the flow velocity vanishes. Figure~\ref{fig:standoff} shows quasi-steady values of $\Delta x_\mathrm{SO}$ after $t\,{=}\,4\hr$ across different simulations, computed using the same procedure as for \Mpdot, $\eta_\mathrm{abl}$, and $\eta_\mathrm{pres}$. With increasing resolution, $\langle\Delta x_\mathrm{SO}\rangle$ generally decreases and is close to convergence for most cases with the $768\times512^2$ grid. Figure~\ref{fig:standoff}(a) confirms the expected decrease of $\langle\Delta x_\mathrm{SO}\rangle$ with \machinf. In contrast, Fig.~\ref{fig:standoff}(b) shows that $\langle\Delta x_\mathrm{SO}\rangle \approx 0.04 \Rsun$, with little variation when $f_\mathrm{m}$ and $f_\mathrm{e}$ are changed at fixed $\machinf=2$. We omitted results from the dense-wind simulations performed in the inertial frame, where the planet's acceleration gradually reduces the velocity contrast with the background flow, leading to a steady increase in $\Delta x_\mathrm{SO}$.

To test the dependence of $\Delta x_\mathrm{SO}$ on the adiabatic index, we performed an additional simulation with $\gamma\,{=}\,7/5$, representative of air, but otherwise identical to the fiducial simulation. This also enables a direct comparison with laboratory wind-tunnel experiments. The results are shown in Fig.~\ref{fig:standoff}(c), where the $\gamma\,{=}\,7/5$ case exhibits significantly smaller $\langle\Delta x_\mathrm{SO}\rangle$ due to higher gas compressibility. At the highest resolution, $\langle\Delta x_\mathrm{SO}\rangle$ agrees within 8\% with an empirical fit to experimental data by \cite{Ambrosio+Wortman62} (orange square marker).

Multiple studies have proposed expressions for the shock stand-off distance based on analytical or numerical models. In Fig.~\ref{fig:standoff}(a) and (c), we include $\Delta x_\mathrm{SO}$ computed using the model by \cite{Kawamura50} and the relation proposed by \cite{Lobb64}. While the \cite{Kawamura50} model captures the relative variation of $\Delta x_\mathrm{SO}$ with \machinf, it also yields systematically smaller values. The simple relation by \cite{Lobb64}, which sets $\Delta x_\mathrm{SO}$ proportional to the density jump across the bow shock, provides the best agreement across the range of \machinf and $\gamma$ tested in our simulations. Recent wind-tunnel hydrodynamical simulations by \cite{Thun+16} and \cite{Prust+Bildsten24} using a rigid sphere without gravity find very close agreement (within a few percent in the latter) with \cite{Ambrosio+Wortman62} for $\gamma\,{=}\,7/5$. This suggests that the deviations in our simulations are likely due to gravitational focusing by the planet, self-gravity of the wake, and the planet's departure from a perfect sphere. For instance, ram-pressure flattening of the planet's front face \citep{Jia+Spruit18} increases the radius of curvature at the stagnation point, leading to a larger shock stand-off distance.

\section{Extra figures}
\label{app:figures}
\begin{figure}
    \centering
    \includegraphics[width=\columnwidth]{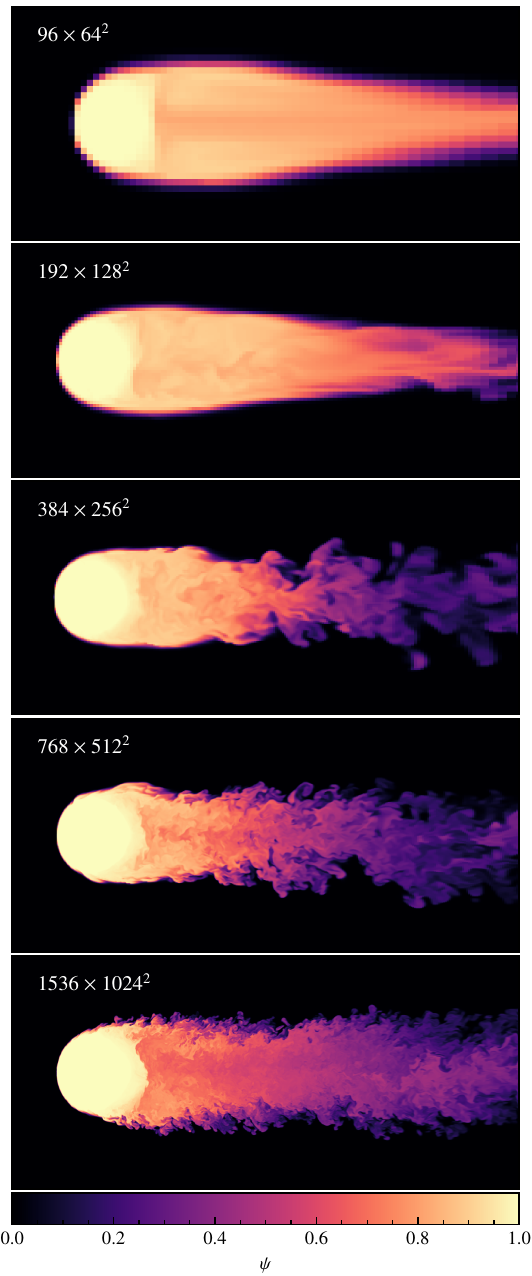}
    \caption{Variation in the flow pattern of the fiducial simulation with increasing resolution. The snapshots show the planet passive scalar, $\psi$, in the $xy$-plane at $t=5\hr$. An animated version of this figure is available via a \href{https://youtube.com/playlist?list=PLZvLCitllq5s5MXv1hH1TnzruyMMslkZ9&si=KW-R2q9AHpAuUZuc}{YouTube playlist}.}
    \label{fig:res}
\end{figure}

\begin{figure}
    \centering
    \includegraphics[width=\columnwidth]{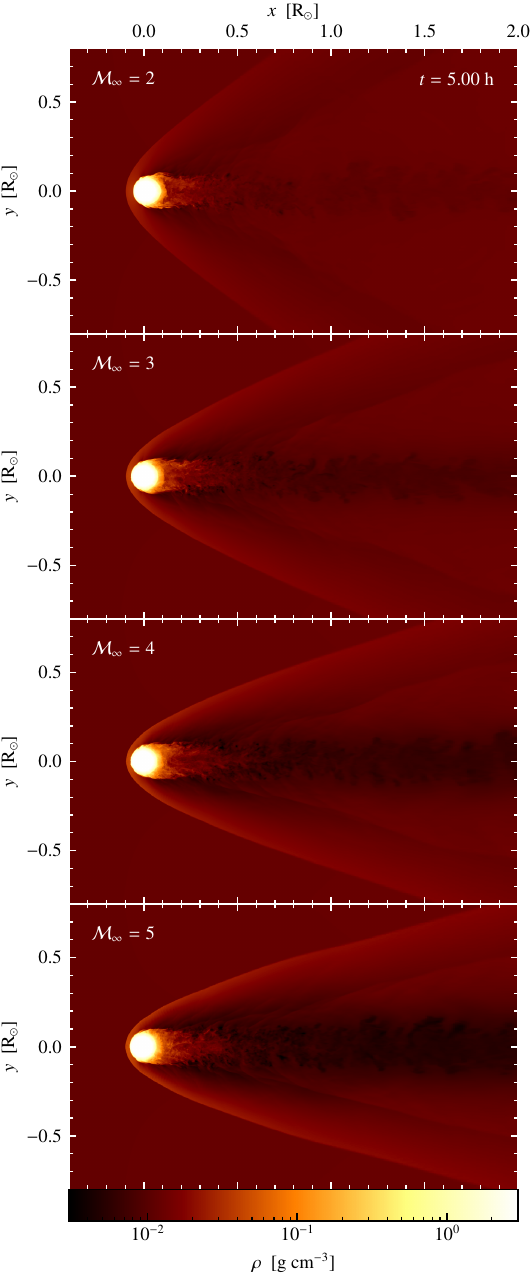}
    \caption{Comparison of simulations with different upstream Mach numbers, \machinf, and with $f_\mathrm{m}\,{=}\,f_\mathrm{e}\,{=}\,1$ at the grid resolution of $768\times512^2$. Each panel shows a density slice in the $xy$-plane at $t=5\hr$. An animated version of this figure is available via a \href{https://youtube.com/playlist?list=PLZvLCitllq5s5MXv1hH1TnzruyMMslkZ9&si=KW-R2q9AHpAuUZuc}{YouTube playlist}.}
    \label{fig:mach-comparison}
\end{figure}

\end{appendix}

\end{document}